\documentclass[%
 reprint,
 amsmath,amssymb,
 aps,
]{revtex4-2}

\usepackage{graphicx}
\usepackage{dcolumn}
\usepackage{bm}

\begin{document}

\preprint{APS/123-QED}

\title{Shape phase transition, coexistence and mixing in the $^{98-106}$Ru isotopes}





\author{R. Budaca$^{1,2}$, P. Buganu$^{1,*}$, F. El Ouardi$^{3}$, A. Lahbas$^{3,4}$}
\address{$^{1}$"Horia Hulubei"$-$National Institute for R\&D in Physics and Nuclear Engineering, Str. Reactorului 30, RO-077125, POB-MG6 Bucharest-M\v{a}gurele, Romania}
\address{$^{2}$ Academy of Romanian Scientists, Splaiul Independentei St., No. 54, Bucharest, P. O. 050094, Romania}
\address{$^{3}$ ESMaR, Department of Physics, Faculty of Science, Mohammed V University in Rabat, Rabat 10000, Morocco}
\address{$^{4}$High Energy Physics and Astrophysics Laboratory, Department of Physics, Faculty of Science Semlalia, Cadi Ayyad University, P. O. B. 2390, Marrakesh 40000, Morocco}

\email[]{buganu@theory.nipne.ro}

\begin{abstract}
The deformation properties within the $^{98-106}$Ru even-even isotopic chain, are investigated by means of the Covariant Density Functional Theory with a Density-Dependent Point-Coupling X parametrization. The considered nuclei are found to exhibit very shallow prolate and triaxial ground state deformation. This information is used to ascertain their dynamical behavior within prolate $\gamma$-stable and $\gamma$-unstable instances of a phenomenological Bohr-Mottelson Hamiltonian with an octic potential in the axial deformation variable. The comparative study of the low-lying collective states, revealed the presence of a shape phase transition from low to high deformation, as well as evidence of shape coexistence and mixing between spherical vibrator, $\gamma$-unstable or prolate configurations in ground and excited states. It is also shown that the effect of shape coexistence and mixing on the $\gamma$-band states can account to some extent for the typical $\gamma$-unstable staggering even in prolate $\gamma$-stable conditions.
\end{abstract}

\maketitle


\section{Introduction}
The nuclear shape phase transitions \cite{Casten,Cejnar} and their critical points \cite{Iachello1,Iachello2} are related to the change in shape of the nucleus in the ground state as a function of the number of nucleons. For example, within an isotopic chain, by varying the number of neutrons one can cover phase transitions \cite{Ginocchio,Dieperink} from a spherical vibrator configuration \cite{Bohr1} of the states to the $\gamma$-unstable \cite{Wilets,Bes} or $\gamma$-stable prolate \cite{Davydov1,Iachello1} and triaxial configuration \cite{Davydov2,Bonatsos1}, respectively. It should be noted that if a nucleus falls in one of these limits or critical points, the excited states of the ground, $\beta$ and $\gamma$ bands preserve the deformation type of the ground state defining in this way specific patterns associated with the corresponding shape.  By contrary, the paradigm changes considerably when the nuclear shape coexistence and mixing phenomena come into play \cite{Heyde1,Heyde2,Bona}. The shape coexistence implies that a low excited state, most often the first excited $0^{+}$ state \cite{Garrett1}, has a distinct shape from the ground state \cite{Morinaga}. Moreover, also the entire bands manifest configurations of the states which correspond to different deformations \cite{Poves}. For example, one can have a spherical vibrator or $\gamma$-unstable configuration of the states for the ground band and a prolate (oblate) configuration for the $\beta$ band. Other combinations of shape coexistence are possible, even triple \cite{Petrovici,Naz} and multiple shape coexistence \cite{Garrett,Yang}. The phenomenon is widespread across the nuclear chart, rather than an isolated one as it was estimated at the beginning. Recently, it has been evidenced the existence of the so-called islands of shape coexistence \cite{Martinou1,Martinou2}. In some situations, states of the same band have different shapes or are characterized by a mixing of shapes. Consequently, one can speak about shape coexistence with and without mixing \cite{BudacaA,Buganu3}.

In the present work, both microscopic and phenomenological models are involved in studying these phenomena. More specifically, the theoretical tools are the Covariant Density Functional Theory (CDFT) with a Density-Dependent Point-Coupling X (DD-PCX) parametrization \cite{Vretenar2005,Niksic2011,PCX_Optimization}, respectively the Bohr-Mottelson Hamiltonian (BMH) \cite{Bohr1,Bohr2} with an octic potential (OP) in the $\beta$ variable for $\gamma$-unstable and prolate cases. The solution of the BMH with OP for $\gamma$-unstable system has been recently proposed  and applied to investigate the shape transition, coexistence and mixing within the Cd isotopic chain \cite{Buganu3}, while the solution for prolate is introduced in the present study. The eigenvalue problem in the prolate case is numerically solved in a basis of Bessel functions of the first kind, following the method given for the sextic potential (SP) \cite{BudacaR,Budaca2}, respectively the solving techniques developed in \cite{Taseli}. This method is different than the quasi-exactly solvable method \cite{Ushveridze} used for the BMH with SP \cite{LevaiArias,Budaca3,Lahbas}. The numerical method offers access to a more general solution than the quasi-exact method making possible to properly address shape coexistence and mixing phenomena and even anomalously small $B(E2)$ transition ratios \cite{Wang,Zhang,Mennana2,Benjedi}. Also, it is worth mentioning that the present numerical method is different than that applied within the General Collective Model \cite{Gneuss,Hess} where solutions of the five-dimensional harmonic oscillator are used as a basis for a general potential in $\beta$ and $\gamma$ with the highest degree in $\beta$ of a sixth order.

Previous applications to the experimental data of the BMH with sextic and octic potentials, involving both numerical \cite{BudacaA,Buganu3,BudacaR,Budaca2,BudacaAP,Benjedi2,Benjedi3} and quasi-exact \cite{LevaiArias,Budaca3,Lahbas} solutions, as well as complementary with the CDFT with Density-Dependent Meson-Exchange (DD-ME2) \cite{Lalazissis} and Density-Dependent Point-Coupling (DD-PC1) \cite{Niksic1} parameter sets \cite{Mennana2,Benjedi,Mennana1,Chafik}, clearly evidence the ability of these approaches to describe such phenomena. Therefore, in the present work a new study is conducted in a concurrent manner for the $^{98-106}$Ru even-even isotopes \cite{Chen,Singh,Frenne1,Blachot,Frenne2}, hoping to contribute to a better understanding of their structure of the states.

The plan of the work is the following. In Section II are presented the theoretical models used in this study, namely the CDFT with DD-PCX parametrization and the BMH with OP for $\gamma$-unstable and prolate deformations, while Section III is dedicated to numerical applications of these models to the experimental data of the $^{98-106}$Ru isotopic chain. A comprehensive discussion on these results, in relation to the corresponding experimental data, is also  made in Section III, while the main achievements of the work are highlighted in Section IV dedicated to conclusions.

\section{Description of the models}

\subsection{Covariant Density Functional Theory with Density-Dependent Point-Coupling interaction}
The microscopic description of nuclear ground-state properties in this work is based on Covariant Density Functional Theory (CDFT) formulated within the relativistic Hartree-Bogoliubov (RHB) framework using the density-dependent point-coupling interaction DD-PCX \cite{Vretenar2005,Niksic2011,PCX_Optimization}.
In this approach, nucleons are described as Dirac particles moving in self-consistent mean fields generated by zero-range four-fermion contact interactions, ensuring Lorentz covariance and an accurate treatment of spin degrees of freedom \cite{Ring2001,Niksic2011}.

\subsubsection{Point-coupling Lagrangian density}

The effective Lagrangian density of the density-dependent point-coupling model is given by
\cite{PCX_Optimization}:
\begin{equation}
\begin{aligned}
\mathcal{L} = &\ \bar{\psi}(i\gamma^\mu \partial_\mu - m)\psi
- \frac{1}{2}\alpha_S(\rho)(\bar{\psi}\psi)(\bar{\psi}\psi)
\\
&- \frac{1}{2}\alpha_V(\rho)(\bar{\psi}\gamma^\mu\psi)(\bar{\psi}\gamma_\mu\psi)\\
&  - \frac{1}{2}\alpha_{TV}(\rho)(\bar{\psi}\vec{\tau}\gamma^\mu\psi)
(\bar{\psi}\vec{\tau}\gamma_\mu\psi)
\\
&- \frac{1}{2}\delta_S(\partial_\nu \bar{\psi}\psi)
(\partial^\nu \bar{\psi}\psi)
- e\bar{\psi}\gamma^\mu A_\mu\frac{1-\tau_3}{2}\psi ,
\end{aligned}
\end{equation}
where $\psi$ denotes the nucleon Dirac field with mass $m$.
The derivative term proportional to $\delta_S$ simulates finite-range effects essential for reproducing nuclear surface properties \cite{Niksic2011}.

The density dependence of the coupling functions is parametrized as \cite{PCX_Optimization}:
\begin{equation}
\alpha_i(\rho) = a_i + (b_i + c_i x)e^{-d_i x},  \qquad (i=S,V,TV),
\end{equation}
where $x=\rho/\rho_{\mathrm{sat}}$, while $\rho_{\mathrm{sat}}$ denotes the saturation density of symmetric nuclear matter. The indices S, V and TV refer to the isoscalar-scalar, isoscalar-vector and isovector-vector interactions between nucleons and $\sigma$, $\rho$ and $\omega$ mesons. In the DD-PCX formulation, the point-coupling CDFT model depends on ten coupling constants governing the scalar, vector, and isovector channels, as reported in Table \ref{tab:DDPCX}.

\subsubsection{Relativistic Hartree--Bogoliubov formalism and pairing correlations}
For open-shell nuclei, pairing correlations are treated self-consistently within the
relativistic Hartree--Bogoliubov (RHB) framework \cite{Vretenar2005,Niksic2011}.
The total energy density functional reads:
\begin{equation}
E_{\mathrm{RHB}}[\rho,\kappa] = E_{\mathrm{RMF}}[\rho] + E_{\mathrm{pair}}[\kappa],
\end{equation}
where $\rho$ is the normal density matrix and $\kappa$ is the pairing tensor.

The pairing energy is given by \cite{Vretenar2005}
\begin{equation}
E_{\mathrm{pair}}[\kappa] =
\frac{1}{4}
\sum_{n_1 n'_1}
\sum_{n_2 n'_2}
\kappa^{*}_{n_1 n'_1}
\langle n_1 n'_1 | V^{PP} | n_2 n'_2 \rangle
\kappa_{n_2 n'_2},
\end{equation}
where $V^{PP}$ denotes the pairing interaction and the indices label the Dirac spinor quantum numbers.

The pairing force is chosen in separable form in momentum space and reads in coordinate space \cite{Tian2009}
\begin{equation}
V^{PP}(\mathbf{r}_1,\mathbf{r}_2,\mathbf{r}'_1,\mathbf{r}'_2)
= -G\,\delta(\mathbf{R}-\mathbf{R}')
P(\mathbf{r})P(\mathbf{r}'),
\end{equation}
where
\begin{equation}
\mathbf{R} = \frac{1}{\sqrt{2}}(\mathbf{r}_1+\mathbf{r}_2),
\qquad
\mathbf{r} = \frac{1}{\sqrt{2}}(\mathbf{r}_1-\mathbf{r}_2),
\end{equation}
and the Gaussian form factor is defined as
\begin{equation}
P(\mathbf{r}) =
\frac{1}{(4\pi a^2)^{3/2}}
\exp\left(-\frac{r^2}{2a^2}\right).
\end{equation}
This pairing interaction preserves translational invariance and provides an accurate description of pairing properties across the nuclear chart \cite{Tian2009}.

\subsubsection{Quadratic constraint method}
The potential energy surfaces are obtained by constrained RHB calculations using the method of quadratic constraints \cite{Niksic2011}.
The total energy functional to be minimized is written as:
\begin{equation}
E' = \langle \hat{H}_{tot} \rangle
+ \sum_{\mu=0,2}
C_{2\mu}
\left(\langle \hat{Q}_{2\mu} \rangle - q_{2\mu}\right)^2 ,
\end{equation}
where $\langle \hat{H}_{tot} \rangle$ is the total energy of the system, $\hat{Q}_{20}=2z^2-x^2-y^2$ and $\hat{Q}_{22}=x^2-y^2$ are the mass quadrupole operators, $q_{2\mu}$ are the constrained values, and $C_{2\mu}$ are stiffness constants.
This procedure allows mapping the potential energy surface as a function of quadrupole deformation parameters $(\beta_2,\gamma)$.

\begin{table}[htbp]
\centering
\caption{Parameters of the density-dependent point-coupling interaction (DD-PCX)
\cite{PCX_Optimization}.
The saturation density is $\rho_{\mathrm{sat}} = 0.152~\mathrm{fm}^{-3}$ and the nucleon
mass is $m = 939~\mathrm{MeV}$.}
\label{tab:DDPCX}
\begin{tabular}{ccc}
\hline
Parameter & Value & Unit \\
\hline
$a_S$ & $-10.979243836$ & fm$^{2}$ \\
$b_S$ & $-9.038250910$  & fm$^{2}$ \\
$c_S$ & $-5.313008820$  & fm$^{2}$ \\
$d_S$ & $1.379087070$   & -- \\
$a_V$ & $6.430144908$   & fm$^{2}$ \\
$b_V$ & $8.870626019$   & fm$^{2}$ \\
$d_V$ & $0.655310525$   & -- \\
$b_{TV}$ & $2.963206854$ & fm$^{2}$ \\
$d_{TV}$ & $1.309801417$ & -- \\
$\delta_S$ & $-0.878850922$ & fm$^{4}$ \\
\hline
$G_n$ & $-800.663126037$ & MeV\,fm$^{3}$ \\
$G_p$ & $-773.776776597$ & MeV\,fm$^{3}$ \\
\hline
\end{tabular}
\end{table}

\subsection{The Bohr-Mottelson Hamiltonian with octic potential}
The phenomenological model Hamiltonian involved in the present study has the expression introduced in \cite{Bohr1,Bohr2}:
\begin{eqnarray}
\hat{H}&=&-\frac{\hbar^{2}}{2B}\left[\frac{1}{\beta^{4}}\frac{\partial}{\partial\beta}\beta^{4}\frac{\partial}{\partial\beta}+\frac{1}{\beta^{2}\sin3\gamma}\frac{\partial}{\partial\gamma}\sin3\gamma\frac{\partial}{\partial\gamma}\right]\nonumber\\
&&+\frac{\hbar^{2}}{8B\beta^{2}}\sum_{k=1}^{3}\frac{\hat{L}_{k}^{2}}{\sin^{2}\left(\gamma-\frac{2\pi}{3}k\right)}+V(\beta,\gamma).
\label{Hamiltonian}
\end{eqnarray}
The notations that appear in Eq. (\ref{Hamiltonian}) are: $B-$mass parameter, $\hbar-$reduced Planck constant, $\hat{L}_{k}-$angular momentum projections in the intrinsic reference frame, $\beta$ and $\gamma$ $-$ intrinsic deformation coordinates, $V(\beta,\gamma)-$energy potential. The model describes quadrupole collective rotations and vibrations of the nuclear surface around an equilibrium shape. A first step in solving the corresponding eigenvalue problem is to consider a separable form of the energy potential \cite{Bohr1,Wilets}:
\begin{equation}
V(\beta,\gamma)=V_{1}(\beta)+\frac{1}{\beta^{2}}V_{2}(\gamma).
\label{potential}
\end{equation}
This selection leads to an exact separation of the $\beta$ degree of freedom from the other four degrees, $\gamma$-deformation and the three rotation Euler angles $(\theta_{1},\theta_{2},\theta_{3})$ \cite{Fortunato1,Buganu1}:
\begin{equation}
\left[-\frac{1}{\beta^{4}}\frac{\partial}{\partial\beta}\beta^{4}\frac{\partial}{\partial\beta}+v_{1}(\beta)-\varepsilon+\frac{\Lambda}{\beta^{2}}\right]F(\beta)=0,
\label{eqbeta}
\end{equation}
\begin{eqnarray}
&&\left[-\frac{1}{\sin3\gamma}\frac{\partial}{\partial\gamma}\sin3\gamma\frac{\partial}{\partial\gamma}+\frac{1}{4}\sum_{k=1}^{3}\frac{\hat{L}_{k}^{2}}{\sin^{2}\left(\gamma-\frac{2\pi}{3}k\right)}\right]\\
&&\times\psi(\gamma,\theta_{i})+v_{2}(\gamma)\psi(\gamma,\theta_{i})=\Lambda\psi(\gamma,\theta_{i}),\;i=1,2,3.\nonumber
\label{eqgamma}
\end{eqnarray}
Here, by $\Lambda$ is denoted the separation constant, while $\varepsilon=\frac{2B}{\hbar^{2}}E$, $v_{1}(\beta)=\frac{2B}{\hbar^{2}}V_{1}(\beta)$ and $v_{2}(\gamma)=\frac{2B}{\hbar^{2}}V_{2}(\gamma)$ are the reduced energy and potentials, respectively. The separation of the $\gamma$ variable from the Euler angles depends on the treatment of the rotational term \cite{Bohr1,Mayer}:
\begin{equation}
\hat{W}=\frac{1}{4}\sum_{k=1}^{3}\frac{\hat{L}_{k}^{2}}{\sin^{2}\left(\gamma-\frac{2\pi}{3}k\right)}.
\end{equation}
Most often, this function is expanded in Taylor series around a minimum of the $\gamma$ potential and kept only the zero order term. There are three main cases, depending on the $\gamma$ deformation conditions, namely: $\gamma$-unstable \cite{Wilets,Bes}, axial-symmetry \cite{Davydov1,Iachello1}, triaxial \cite{Davydov2}. More specific cases may also arise if for example a $\gamma$-rigidity is considered \cite{Bonatsos1,Bonatsos2}. In the present study, only the $\gamma$-unstable and axially-deformed symmetries are involved. The solutions for these two cases are presented in the next subsections, after the solution of the $\beta$ equation with octic potential is firstly addressed.

\subsubsection{Solution of the $\beta$-equation}
In what follows, it is more convenient to expand the kinetic term in $\beta$ of Eq. (\ref{eqbeta}) so that:
\begin{equation}
\left[-\frac{d^{2}}{d\beta^{2}}-\frac{4}{\beta}\frac{d}{d\beta}+v_{1}(\beta)-\varepsilon+\frac{\Lambda}{\beta^{2}}\right]F(\beta)=0.
\label{eqredbeta}
\end{equation}
Further, this equation is solved for an octic potential of the form:
\begin{equation}
v_{1}(\beta)=\beta^{2}+b_{1}\beta^{4}+b_{2}\beta^{6}+b_{3}\beta^{8},
\label{OP}
\end{equation}
where $b_{1}$, $b_{2}$ and $b_{3}$  are free parameters. The $\beta^{2}$ term of the potential has no parameter due to a scaling property of the polynomial potentials \cite{Budaca1,Buganu2}. Eq. (\ref{eqredbeta}) with the octic potential (\ref{OP}) is numerically solved here in a basis of the Bessel functions of the first kind \cite{Taseli}. These functions are solutions of Eq. (\ref{eqredbeta}) for an Infinite Square Well potential (ISW) \cite{Iachello1,Iachello2}:
\begin{eqnarray}
	v_{ISW}(\beta)=\bigg \{
	  \begin{matrix}
	  0, & \beta\leq\beta_{\omega}, \\
	  \infty, & \beta>\beta_{\omega}. \\
	  \end{matrix}
\label{iswp}
\end{eqnarray}
Here, $\beta_{\omega}$ is the value of $\beta$ in which the potential becomes infinite. The corresponding equation in this case becomes:
\begin{equation}
\left(-\frac{d^{2}}{d\beta^{2}}-\frac{4}{\beta}\frac{d}{d\beta}-\varepsilon_{ISW}+\frac{\Lambda}{\beta^{2}}\right)f(\beta)=0,
\end{equation}
where by $\varepsilon_{ISW}$ is denoted the energy eigenvalue. In order to reach a Bessel-type form, the following change of function and variable, respectively notations are adopted:
\begin{equation}
f(\beta)=\beta^{-\frac{3}{3}}J_{\nu}(\beta),\, z=\beta k,\, \varepsilon_{ISW}=k^{2},\, \nu=\sqrt{\Lambda+\frac{9}{4}},
\end{equation}
\begin{equation}
\left[\frac{d^{2}}{dz^{2}}+\frac{1}{z}\frac{d}{dz}+\left(1-\frac{\nu^{2}}{z^{2}}\right)\right]J_{\nu}(z)=0.
\label{Bessel}
\end{equation}
The eigenvalues, respectively the eigenfunctions of the Bessel equation (\ref{Bessel}), are obtained from the boundary condition
$J_{\nu}(\beta_{\omega})=0$ imposed for the Bessel function of the first kind of index $\nu$:
\begin{equation}
\varepsilon_{ISW;n,\nu}=\frac{z_{n,\nu}^{2}}{\beta_{\omega}^{2}},\,f_{n,\nu}(\beta)=\frac{\sqrt{2}}{\beta_{\omega}}\frac{\beta^{-\frac{3}{2}}J_{n,\nu}\left(\frac{z_{n,\nu}}{\beta_{\omega}}\beta\right)}{J_{\nu+1}(z_{n,\nu})}.
\end{equation}
Here, $z_{n,\nu}$ is the $n^{th}$ Bessel zero of index $\nu$.
Having now the basis determined, one can further construct the wave functions for the octic potential such as:
\begin{equation}
F_{\xi,\nu}(\beta)=\sum_{n=1}^{n_{Max}}A_{n}^{\xi}f_{n,\nu}(\beta),
\label{totwav}
\end{equation}
where $n_{Max}$ gives the dimension of the basis, $\xi$ is related to the $\beta$-vibration number $n_{\beta}$ through the relation $\xi=n_{\beta}+1$, while by $A_{n}^{\xi}$ are denoted the eigenvector components. The general matrix element is easily obtained by taking the advantage of knowing the eigenvalue for the ISWP and by calculating the matrix elements $I_{mn}^{(\nu,i)}$ of the octic potential terms:
\begin{equation}
H_{mn}=\left(\frac{z_{n,\nu}}{\beta_{\omega}}\right)^{2}\delta_{mn}+\beta_{\omega}^{2}I_{mn}^{(\nu,1)}+\sum_{i=2}^{4}b_{i-1}\beta_{\omega}^{2i}I_{mn}^{(\nu,i)}.
\label{matrixelements}
\end{equation}
It is preferable to introduce the following change of variable $x=\beta/\beta_{\omega}$ for the matrix elements over $\beta$:
\begin{eqnarray}
I_{mn}^{(\nu,i)}&=&\frac{2}{J_{\nu+1}(z_{m,\nu})J_{\nu+1}(z_{n,\nu})}\times\nonumber\\
&&\int_{0}^{1}J_{\nu}(z_{m,\nu}x)J_{\nu}(z_{n,\nu}x)x^{2i+1}dx,
\end{eqnarray}
which extract out the dependence on $\beta_{\omega}$. This property assures an independence of any parameter for these matrix elements, which is useful in numerical calculations because they can be determined only once. In order to complete the solvability of the equation in the $\beta$ variable, one needs to find the expression of the index $\nu$ of the Bessel function, which in turn depends on the eigenvalue $\Lambda$ of the equation in the $\gamma$ variable and the three Euler angles. In what follows, the equation is solved for the $\gamma$-unstable and prolate symmetries.

\subsubsection{Solution of the $\gamma$-equation in the $\gamma$-unstable case}
This solution has been already proposed and applied to study the shape coexistence and mixing in the $^{106-116}$Cd isotopes \cite{Buganu3}. Here, it will be briefly presented and applied to new experimental data.
Thus, the energy potential in this case is $\gamma$-independent leading to the eigenvalue problem of the Casimir operator $\hat{\Lambda}$ of the SO(5) group whose eigenvalue is given by \cite{Bohr1,Bes,Rakavy}:
\begin{equation}
\Lambda=\tau(\tau+3),\;\tau=0,1,2,...
\end{equation}
where, $\tau$ is called the seniority quantum number. The separation constant determines the index $\nu$ of the Bessel function,
\begin{equation}
\nu=\sqrt{\tau(\tau+3)+\frac{9}{4}}=\tau+\frac{3}{2},
\end{equation}
and further the total energy of the $\gamma$-unstable system for the octic potential:
\begin{equation}
E_{\xi,\tau,L}=\frac{\hbar^{2}}{2B}\left[\varepsilon_{\xi,\tau}(b_{1},b_{2},b_{3})+cL(L+1)\right].
\label{eunstable}
\end{equation}
The last term, $L(L+1)$, is the eigenvalue of the $SO(3)$ symmetry operator $\hat{L}^{2}$ and it was added to broke the degeneracy over $\tau$ \cite{Caprio}. For numerical applications to the experimental data it is more appropriate to use the following normalization of the energies:
\begin{equation}
R_{\xi,\tau,L}=\frac{E_{\xi,\tau,L}-E_{1,0,0}}{E_{1,1,2}-E_{1,0,0}}.
\label{norunst}
\end{equation}
This allows us to determine the free parameters $b_{1}$, $b_{2}$, $b_{3}$ and $c$ independently on the scaling parameter $\frac{\hbar^{2}}{2B}$. This scaling parameter is fixed at the end such that to exactly reproduce the absolute experimental value of the first excited energy of the ground band. Concerning the parameter $\beta_{\omega}$, defining the ISWP, this is determined from the condition of its tangential intersection of the octic potential tail. Thus, $\beta_{\omega}$ is expressed as a function of the free parameters of the octic potential. In order to establish the nature of each state, of which band belong, a correlation between the energy levels and quadrupole $E2$ electromagnetic transitions is mandatory. These transitions are calculated with the following operator \cite{Wilets}:
\begin{eqnarray}
T_{2,\mu}^{(E2)}&=&t\beta D_{\mu,2}^{(2)}(\theta_{i})\cos\gamma+\nonumber\\
&&\frac{t\beta}{\sqrt{2}}\left[D_{\mu,2}^{(2)}(\theta_{i})+D_{\mu,-2}^{(2)}(\theta_{i})\right]\sin\gamma,
\label{tranop}
\end{eqnarray}
where $t=3R^{2}Ze\beta_{M}/4\pi$ depends on the nuclear radius $R=R_{0}A^{1/3}$, charge number $Z$, elementary charge $e$ and a scaling parameter $\beta_M$. Here, $R_{0}=1.2$ fm, $A$ is the atomic mass, while by $D(\theta_{i})$ denote the Wigner matrices \cite{Wigner}. Also, the expression for the $E2$ transitions incorporates the contribution of all five variables:
\begin{equation}
\Psi_{\xi,\tau,L,M,K}^{(unstable)}(\beta,\gamma,\theta_{i})=F_{\xi,\tau}(\beta)\psi_{\tau,L,M,K}(\gamma,\theta_{i}),
\end{equation}
where the additional quantum numbers $L$, $M$ and $K$ define the eigenvalues of the total angular momentum and its projections on the third body fixed axis and the laboratory z-axis, respectively.
Thus, the $B(E2)$ electromagnetic transitions are given by:
\begin{eqnarray}
&&B(E2;\xi\tau L\rightarrow\xi'\tau' L')=\left(\frac{3R^{2}Ze}{4\pi}\right)^{2}\beta_{M}^{2}\times\nonumber\\
&&(\tau'\alpha'L';112||\tau\alpha L)^{2}\langle\tau|||Q|||\tau'\rangle^{2}(B_{\xi\tau;\xi'\tau'})^{2},
\label{BE2}
\end{eqnarray}
where, $(\tau_{1}\alpha_{1}L_{1};\tau_{2}\alpha_{2}L_{2}||\tau_{3}\alpha_{3}L_{3})$ is the $SO(5)$ Clebsch-Gordon coefficient \cite{Rowe1}, while
\begin{equation}
\langle\tau|||Q|||\tau'\rangle=\sqrt{\frac{\tau}{2\tau+3}}\delta_{\tau,\tau'+1}+\sqrt{\frac{\tau+3}{2\tau+3}}\delta_{\tau,\tau'-1},
\label{tauop}
\end{equation}
is the $SO(5)$ reduced matrix element of the quadrupole moment \cite{Rowe2,Rowe3}.
The integral over $\beta$ is denoted here by:
\begin{equation}
B_{\xi\tau;\xi'\tau'}=\langle F_{\xi,\tau}(\beta)|\beta|F_{\xi',\tau'}(\beta)\rangle.
\end{equation}
Another relevant quantity for the present study, is the probability density distribution of deformation in $\beta$:
\begin{equation}
\rho_{\xi,\tau}(\beta)=[F_{\xi,\tau}(\beta)]^{2}\beta^{4}.
\label{density}
\end{equation}
This is evaluated for a better interpretation of the experimental data in relation to the effective octic potential
\begin{equation}
v_{eff}(\beta)=\frac{\Lambda+2}{\beta^{2}}+\beta^{2}+b_{1}\beta^{4}+b_{2}\beta^{6}+b_{3}\beta^{8},
\label{effpot}
\end{equation}
respectively to the energy levels given by Eq. (\ref{eunstable}). Correlations between these quantities and the monopole transition $E0$ \cite{Wood},
\begin{equation}
\rho^{2}(E0;0_{2}^{+}\rightarrow0_{1}^{+})=\left(\frac{3Z}{4\pi}\right)^{2}\beta_{M}^{4}\langle F_{2,0}(\beta)|\beta^{2}|F_{1,0}(\beta)\rangle^{2},
\label{monopole}
\end{equation}
are made in order to establish the presence or absence of the shape coexistence and mixing phenomena.

\subsubsection{Solution of the $\gamma$-equation in the $\gamma$-stable prolate case}
This solution, of the Bohr-Mottelson Hamiltonian with octic potential in the $\beta$ variable for a prolate deformation, represents an original contribution of the present study. The main difference between these two solutions, $\gamma$-unstable and $\gamma$-stable (prolate), consists in the treatment of the $\gamma$ variable. In the first case, the deformation in $\gamma$ is not stable, the $\gamma$ wave function  oscillating between $\gamma_{0}=0^{o}$ (prolate) and $\gamma_{0}=60^{o}$ (oblate) with a extended maximum centered around $\gamma_{0}=30^{o}$ (triaxial) \cite{Fortunato1}. This situation is also known in literature as the $\gamma-$soft case \cite{Casten}, but sometimes this terminology refers to the presence of the $\gamma$ vibrations as well as the $\gamma$-rigid concept means the opposite, namely the absence of the $\gamma$-vibrations. Thus, here by $\gamma-$stable prolate one understands that the general potential (\ref{potential}) has an explicit dependence on the $\gamma$ variable, respectively $v_{2}(\gamma)$ has a localized minimum around $\gamma_{0}=0^{o}$. In the present study, it is adopted the form of the $\gamma$-potential introduced in \cite{Budaca2}:
\begin{equation}
v_{2}(\gamma)=2q^{2}(1-\cos3\gamma)+d,
\end{equation}
where $q$ and $d$ are free parameters. The separation of the $\gamma$ variable from the three Euler angles $\theta_{i}$ is achieved by considering the wave function $\psi(\gamma,\theta_{i})=\eta(\gamma)D_{M,K}^{(L)}(\theta_{i})$ and the zero order approximation around $\gamma_{0}=0^{o}$ for the rotational energy:
\begin{equation}
\frac{1}{4}\sum_{k=1}^{3}\frac{\hat{L}_{k}^{2}}{\sin^{2}\left(\gamma-\frac{2\pi}{3}k\right)}\approx\frac{\hat{L}^{2}}{3}+\hat{L}_{3}^{2}\left(\frac{1}{4\sin^{2}\gamma}-\frac{1}{3}\right).
\label{opomega}
\end{equation}
Here, $D_{M,K}^{(L)}(\theta_{i})$ are the rotational Wigner functions \cite{Wigner}, which are eigenstates for the above rotational operator, with the indices $L$, $M$ and $K$ denoting quantum numbers for the eigenvalues of the total angular momentum and its projections on the third body fixed axis and the laboratory z-axis, respectively. The action of the operator, given by Eq. (\ref{opomega}), on the Wigner functions $D_{M,K}^{(L)}(\theta_{i})$ gives:
\begin{equation}
\frac{L(L+1)-K^{2}}{3}+\frac{K^{2}}{\sin^{2}\gamma},
\end{equation}
which leads to an equation only in the $\gamma$ variable. Then, in the small angle approximation $\gamma\approx0^{o}$, the $\gamma$ equation is reduced to a differential equation for a two-dimensional harmonic oscillator \cite{Iachello1}:
\begin{equation}
\left[-\frac{1}{\gamma}\frac{\partial}{\partial\gamma}\gamma\frac{\partial}{\partial\gamma}+\frac{K^{2}}{4\gamma^{2}}+(3q)^{2}\gamma^{2}\right]\eta(\gamma)=\epsilon_{\gamma}\eta(\gamma),
\end{equation}
where \cite{Budaca2},
\begin{equation}
\epsilon_{\gamma}=6q(n_{\gamma}+1)=\Lambda-\frac{L(L+1)-K^{2}}{3}-d
\end{equation}
Here, $n_{\gamma}$ is the quantum number associated with the vibrations in the $\gamma$ variable and its values are determined from the following relation:
\begin{equation}
n_{\gamma}=2n+\frac{|K|}{2},\;n=0,1,2,...
\end{equation}
Further, by considering $d=-6q$ as in \cite{Budaca2}, one gets the expression for the separation constant $\Lambda$:
\begin{equation}
\Lambda=\frac{L(L+1)-K^2}{3}+6qn_{\gamma}.
\label{ompro}
\end{equation}
This allow us to calculate the index $\nu$ of the Bessel functions:
\begin{equation}
\nu=\sqrt{\Lambda+\frac{9}{4}}=\sqrt{\frac{L(L+1)-K^2}{3}+6qn_{\gamma}+\frac{9}{4}},
\end{equation}
respectively the energies:
\begin{equation}
E_{\xi,n_{\gamma},L,K}=\frac{\hbar^{2}}{2B}\varepsilon_{\xi,n_{\gamma},L,K}(b_{1},b_{2},b_{3},q),
\end{equation}
and of the wave functions (\ref{totwav}). Energies can be also normalized to the energy of the first excited state of the ground band:
\begin{equation}
R_{\xi,n_{\gamma},L,K}=\frac{E_{\xi,n_{\gamma},L,K}-E_{1,0,0,0}}{E_{1,0,2,0}-E_{1,0,0,0}}.
\label{norpro}
\end{equation}
Reassembling the total wave function, taking into account all five variables, one obtains:
\begin{equation}
\Psi_{\xi,n_{\gamma},L,K}^{(stable)}(\beta,\gamma,\theta_{i})=F_{\xi,\nu}(\beta)\eta_{n_{\gamma},L,K}(\gamma)\Theta_{L,M,K}(\theta_{i}),
\end{equation}
where,
\begin{eqnarray}
\Theta_{L,M,K}(\theta_{i})&=&\sqrt{\frac{2L+1}{16\pi^{2}(1+\delta_{K,0})}}\times\nonumber\\
&&\left[D_{M,K}^{(L)}(\theta_{i})+(-1)^{L}D_{M,-K}^{(L)}(\theta_{i})\right].
\end{eqnarray}
Having the wave functions, one can further calculate the $B(E2)$ transitions using the transition operator given by Eq. (\ref{tranop}):
\begin{eqnarray}
&&B(E2;\xi,n_{\gamma}LK\rightarrow\xi'n_{\gamma}'L'K')=\left(\frac{3R^{2}Ze}{4\pi}\right)^{2}\beta_{M}^{2}\times\nonumber\\
&&\left(B_{\xi,n_{\gamma}LK}^{\xi'n_{\gamma}'L'K'}G_{n_{\gamma}K}^{n_{\gamma}'K'}C_{KK'-KK'}^{L2L'}\right)^{2}.
\end{eqnarray}
The constants in front of the above equation are defined in the previous subsection. Here, by $B$ and $G$ are denoted the contributions from the $\beta$ and $\gamma$ variables, while by $C$ the Clebsch-Gordon coefficient resulted from the matrix elements over the Euler angles. In the small angle approximation, $G$ takes the values $1$ for $\triangle K=0$, respectively $1/\sqrt{3q}$ for $|\triangle K=2|$. The monopole $E0$ transition and the probability density distribution of deformation are calculated in the same way as in the previous subsection, excepting that here one has a different expression for the index of the wave function. For example, $\rho_{\xi,\tau}(\beta)$ (\ref{density}) is amended with the $K$ quantum number $\rho_{\xi,K,\tau}(\beta)$, the latter being $K=0$ for the ground and $\beta$ bands, respectively $K=2$ for the $\gamma$ band. Also, the effective potential in this case is simply obtained just by replacing Eq. (\ref{ompro}) in Eq. (\ref{effpot}). Also, in the next section dedicated for numerical results and applications to the experimental data, it is introduced the variable $\beta_{2}=\beta_{M}\beta$ defining the magnitude of the quadrupole deformation with the traditional microscopic domain of values.

\section{Numerical results}
The $^{98-106}$Ru even-even isotopes are selected for numerical applications of the models discussed in the previous section. The isotopes of Ru have been previously investigated with different approaches for shape phase transitions and critical points, respectively for the presence of the shape coexistence and mixing phenomena. For example, a quasi-exactly solvable method \cite{Ushveridze} of the Bohr-Mottelson Hamiltonian with sextic oscillator potential has been involved in finding the best Ru isotope candidate for the critical point of the phase transitions between the spherical vibrator and prolate rotor  \cite{Buganu4}, respectively $\gamma$-unstable system \cite{Buganu4,Levai,Budaca3,Lahbas}. Also, the Generalized Collective Model \cite{Gneuss,Hess}, with a potential containing mixed terms in $\beta$ and $\gamma$ variables and going up to a sixth order polynomial in $\beta$,  was numerically solved and applied much earlier to this isotope chain in order to describe both shape phase transitions and coexistence phenomena \cite{Troltenier}. Results for shape evolution and ground state properties obtained with the CDFT for these isotopes using DD-ME2  and DD-PC1 parametrizations \cite{Niksic1,Niksic2,Lalazissis} can be found in \cite{Bassem}, and at some level even the new DD-PCX parametrization restricted only to the axial configurations has also been applied \cite{Thakur}. Of great reference for these topics are calculations made with the Interacting Boson Model \cite{Iachello3} with configuration mixing \cite{Barbecho}, with a microscopic input based on a Gogny energy density functional \cite{Nomura}, or making use of $d-$boson occupation and quadrupole interaction probabilities \cite{Hosseinnezhad}, respectively involving the coherent state formalism \cite{Ginocchio,Dieperink} in \cite{Khalaf}. In these latter studies, a special attention was given to the influence of the shape coexistence on the structure of the states. The list of applications can continue with the relativistic-Hartree-Bogoliubov formalism using density-dependent zero and finite range $NN$ interactions and with separable pairing \cite{Abusara} evidencing triaxiality softness in all the isotopes of Ru, respectively beyond-mean-field calculations \cite{Garrett1,Garrett2} supporting the presence of different degrees of coexistence and mixing configurations between spherical, triaxial or prolate shapes for these isotopes. Thus, the present models, namely the Bohr-Mottelson Hamiltonian with octic potential for $\gamma$-unstable and prolate deformations come with a new approach of treating this isotopic chain, hoping to bring new progress in understanding their structure. For example, the former solution can describe phase transitions, coexistence and mixing between an approximately spherical shape and a $\gamma$-unstable deformed one, while the latter solution between an approximately spherical shape and a prolate one. Moreover, when the two solutions are applied together, which is the case here, coexistence and mixing between triaxial ($\gamma$-unstable) and prolate shapes can be also captured. Complementary, the CDFT with DD-PCX is involved here in studying the ground state properties of these isotopes, offering a microscopic support for the two phenomenological approaches.

\begin{figure}
	\centering
\includegraphics[width=0.4\textwidth]{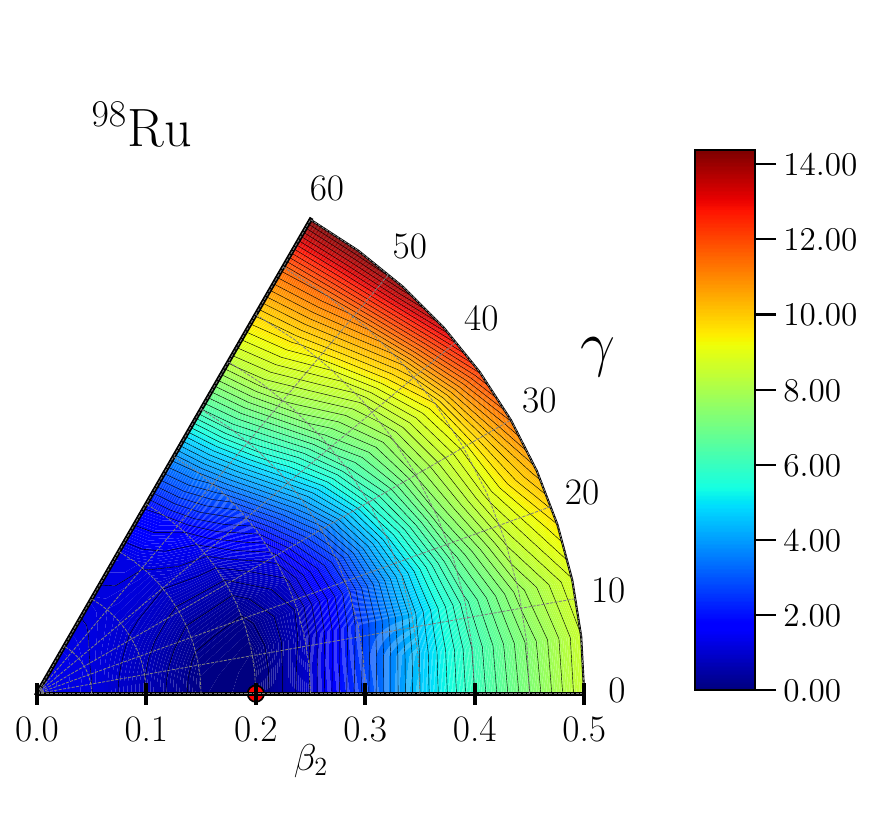}
\includegraphics[width=0.4\textwidth]{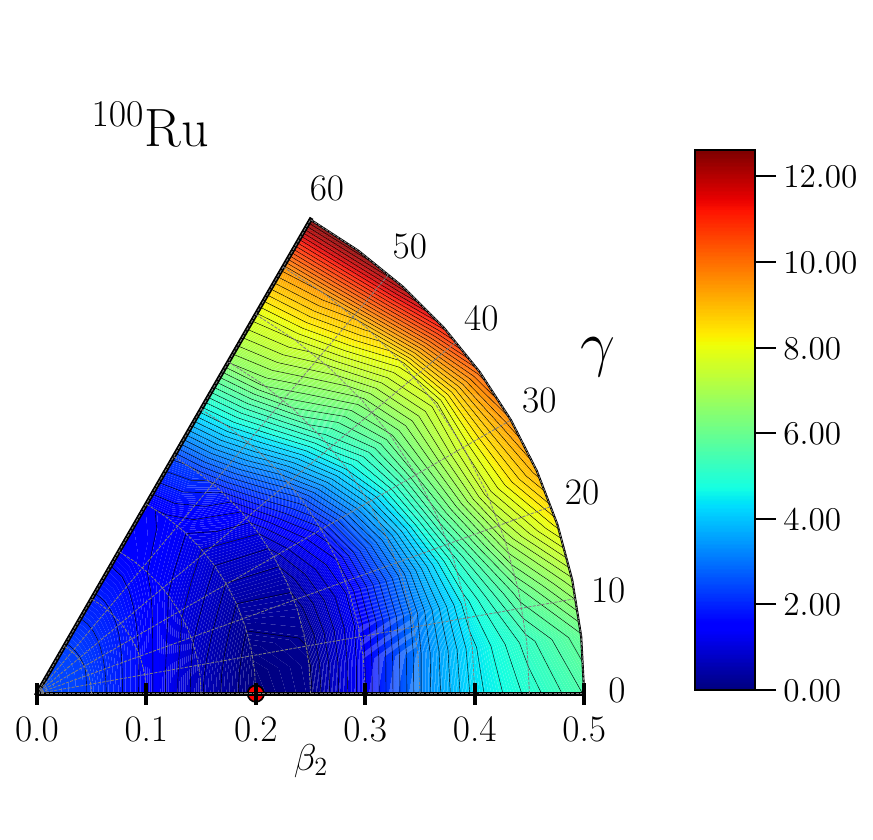}
	\caption{(Color online) Potential energy surfaces in the $(\beta_{2},\gamma)$ plane for $^{98}$Ru and $^{100}$Ru obtained from the CDFT calculations with DD-PCX interaction. The red dot indicates the minimum point, in these cases corresponding to prolate shapes, while the color scale, given in MeV units, shows the relative energy potential depth.}
	\label{fig1}
\end{figure}

\begin{figure}
	\centering
\includegraphics[width=0.4\textwidth]{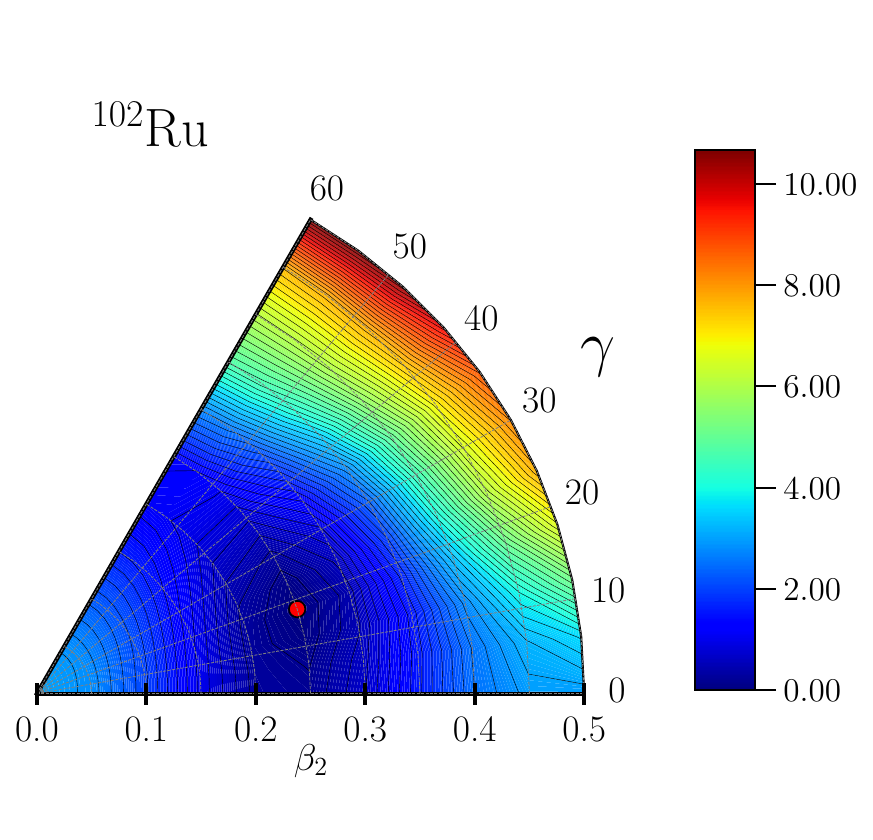}
\includegraphics[width=0.4\textwidth]{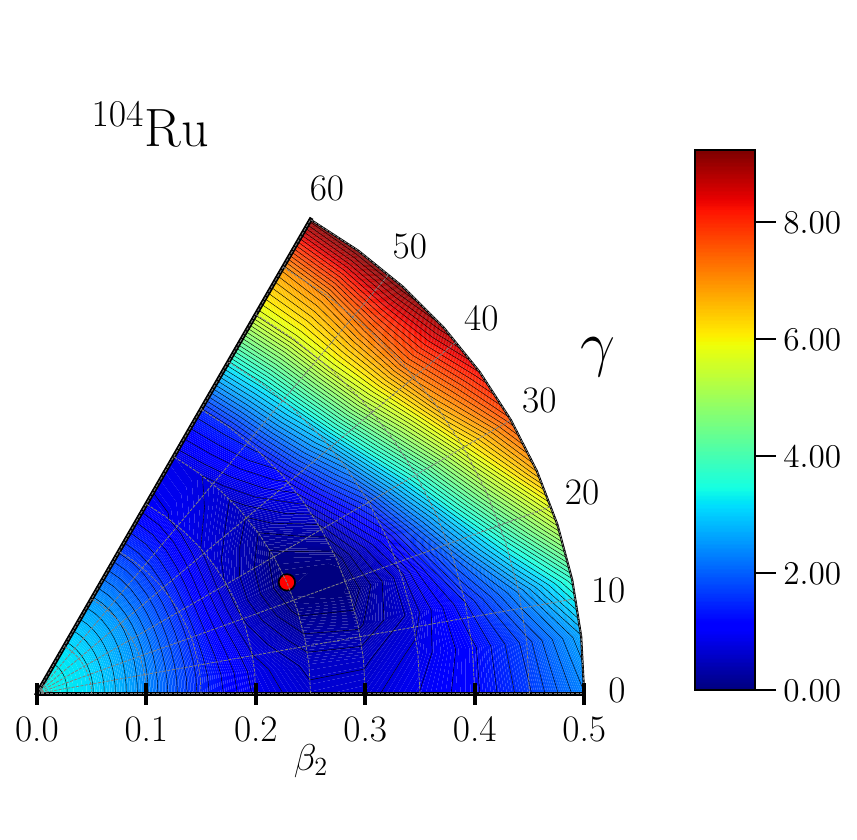}
\includegraphics[width=0.4\textwidth]{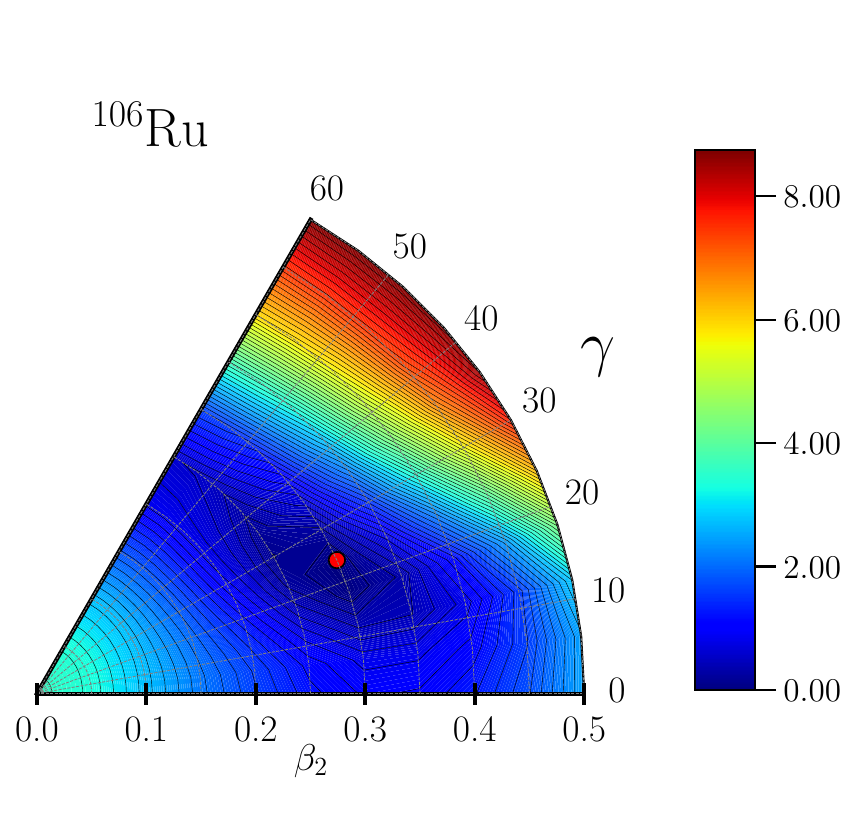}
	\caption{(Color online) Potential energy surfaces in the $(\beta_{2},\gamma)$ plane for $^{102}$Ru, $^{104}$Ru and $^{106}$Ru obtained from the CDFT calculations with DD-PCX interaction. The red dot indicates the minimum point, in these cases corresponding to triaxial shapes, while the color scale, given in MeV units, shows the relative energy potential depth. }
	\label{fig2}
\end{figure}

Therefore, in Figs. \ref{fig1} and \ref{fig2} are plotted the potential energy surfaces obtained with the CDFT with DD-PCX for each isotope. The ground state deformations, indicated by a red dot on the figures, are extracted and compared in Table \ref{tab2} with the corresponding values calculated with the Bohr-Mottelson Hamiltonian (BMH) for both $\gamma$-unstable and prolate cases as well as by those given in the database of ENSDF \cite{beta2}. Concerning the $\beta_{2}$ deformation, one can see that there is a good agreement between the CDFT and the BMH predictions, respectively a partial one with the ENSDF data. More specifically, with the latter ones there is a very good match for $^{98}$Ru, $^{100}$Ru and $^{104}$Ru, and less good, but still close for $^{102}$Ru and $^{106}$Ru. According to CDFT, one has a prolate deformation in the ground state for $^{98,100}$Ru and an increasing triaxial deformation for the heavier $^{102,104,106}$Ru isotopes. The $\gamma$ values given in Table \ref{tab2} for BMH are considered $\gamma=0^{o}$, the minimum of the $\gamma$ potential, and $\gamma=30^{o}$, which is the maximum of the $\gamma$-wave function in the $\gamma$-unstable case. All these results are important being of reference in the next step of the investigation of collective excited states of these isotopes. A special attention is accorded to the lowest quadrupole collective states of the ground, $\beta$ and $\gamma$ bands, where phenomena such as dynamical shape evolution, shape coexistence and mixing are more expected to appear.

\begin{table}
\caption{The $(\beta_{2},\gamma)$ ground state deformations, calculated with CDFT and BMH for the even-even $^{98-106}$Ru isotopes, are presented together with the corresponding $\beta_{2}$ values taken from ENSDF \cite{beta2}}
\begin{center}
\resizebox{8.5cm}{1.1cm} {
\begin{tabular}{c|ccccc}
\hline
\hline
$(\beta_{2},\gamma)$&$^{98}$Ru&$^{100}$Ru&$^{102}$Ru&$^{104}$Ru&$^{106}$Ru\\
\hline
\hline
CDFT             &$(0.200,0^{o})$&$(0.200,0^{o})$&$(0.250,18^{o})$&$(0.250,24^{o})$&$(0.300,24^{o})$\\
$\gamma$-unstable&$(0.187,30^{o})$&$(0.208,30^{o})$&$(0.237,30^{o})$&$(0.267,30^{o})$&$(0.288,30^{o})$\\
$\gamma$-prolate &$(0.186,0^{o})$&$(0.200,0^{o})$&$(0.234,0^{o})$&$(0.259,0^{o})$&$(0.281,0^{o})$\\
ENSDF            &$(0.197,$-$)$&$(0.215,$-$)$&$(0.171,$-$)$&$(0.274,$-$)$&$(0.207,$-$)$\\
\hline
\hline
\end{tabular}}
\end{center}
\label{tab2}
\end{table}

In Tables \ref{tab3} and \ref{tab4} are given the parameters of the two approaches of the Bohr-Mottelson Hamiltonian fitted for the experimental data of the $^{98-106}$Ru isotopes \cite{Chen,Singh,Frenne1,Blachot,Frenne2,Garrett1}. The free parameters $b_{1}$, $b_{2}$, $b_{3}$ of the octic potential plus an additional one, $c$ for the $\gamma$-unstable case and $q$ for the prolate case, are fitted for the normalized energies given by Eqs. (\ref{norunst}) and (\ref{norpro}) involving the squares mean root procedure ($rms$):
\begin{equation}
rms=\sqrt{\frac{\sum_{N}(R_{N}^{Theory}-R_{N}^{Experiment})^{2}}{N}},
\end{equation}
where by $R$ is denoted the normalized energy, while $N$ represents the number of the experimental states included in the fit.
Concerning the scaling factors $\hbar^{2}/2B$ and $\beta_{M}$, these are fixed to reproduce the experimental energy of the first excited state of the ground band and the $B(E2)$ electromagnetic transition between the first excited and ground states of the ground band, respectively. All these free and scaling parameters are then used to calculate the energy spectrum, the $E2$ and $E0$ electromagnetic transitions, respectively to plot the effective energy potentials and the probability density distributions of deformation. These quantities and results are shown in Figures \ref{fig3}$-$\ref{fig22}, being further analyzed and discussed for a better understanding of the structure of the states for these isotopes.
\begin{table}
\caption{Parameters of the Bohr-Mottelson Hamiltonian with octic potential for $\gamma$-unstable deformation (Octic $\&$ $\gamma$-unstable) fitted through the root mean square procedure (rms) for the experimental data of the $^{98-106}$Ru isotopes \cite{Chen,Singh,Frenne1,Blachot,Frenne2,Garrett1}. Starting with the lighter isotope, the parameters $b_{1}$, $b_{2}$ and $b_{3}$ lead to the following values for $\beta_\omega$: 2.4, 1.5, 1.6, 1.7, 1.8. }
\begin{center}
\resizebox{8.5cm}{1.4cm} {
\begin{tabular}{cccccccc}
\hline
$\gamma$-unstable&\multicolumn{4}{c}{Free parameters}&&\multicolumn{2}{c}{Scaling factors}\\
\cline{2-5}\cline{7-8}
Isotope&$b_{1}$&$b_{2}$&$b_{3}$&$c$&$rms$&$\frac{\hbar^{2}}{2B}$ [keV]&$\beta_{M}$\\
\cline{2-5}\cline{7-8}
$^{98}$Ru&21.44&$-$9.61&1.22&0.167&0.148&71.09&0.251\\
$^{100}$Ru&$-$28.56&49.94&0.101&0.156&0.315&46.99&0.316\\
$^{102}$Ru&6.84&49.99&0.100&0.601&0.418&27.57&0.408\\
$^{104}$Ru&$-$46.16&38.88&0.100&0.258&0.314&41.08&0.335\\
$^{106}$Ru&$-$41.49&21.93&0.101&0.239&0.558&46.20&0.287\\
\hline
\end{tabular}}
\end{center}
\label{tab3}
\end{table}

\begin{table}
\caption{Parameters of the Bohr-Mottelson Hamiltonian with octic potential for prolate deformation (Octic $\&$ prolate) are fitted through the root mean square procedure (rms) for the experimental data of the $^{98-106}$Ru isotopes \cite{Chen,Singh,Frenne1,Blachot,Frenne2,Garrett1}. Starting with the lighter isotope, the parameters $b_{1}$, $b_{2}$ and $b_{3}$ lead to the following values for $\beta_\omega$: 3.1, 2.4, 4.0, 2.2, 2.8.  }
\begin{center}
\resizebox{8.5cm}{1.4cm} {
\begin{tabular}{cccccccc}
\hline
$\gamma$-prolate&\multicolumn{4}{c}{Free parameters}&&\multicolumn{2}{c}{Scaling factors}\\
\cline{2-5}\cline{7-8}
Isotope&$b_{1}$&$b_{2}$&$b_{3}$&$q$&$rms$&$\frac{\hbar^{2}}{2B}$ [keV]&$\beta_{M}$\\
\cline{2-5}\cline{7-8}
$^{98}$Ru &5.00&$-$1.70&0.148&1.07&0.236&268.1&0.182\\
$^{100}$Ru&16.74&$-$9.21&1.27&1.51&0.190&274.1&0.177\\
$^{102}$Ru&0.546&$-$0.115&0.006&1.35&0.412&337.1&0.174\\
$^{104}$Ru&19.33&$-$12.40&2.00&1.36&0.236&173.3&0.233\\
$^{106}$Ru&9.75&$-$6.20&1.00&1.54&0.301&146.75&0.239\\
\hline
\end{tabular}}
\end{center}
\label{tab4}
\end{table}

Analyzing the $rms$ values, calculated in Tables \ref{tab3} and \ref{tab4} with the Bohr-Mottelson Hamiltonian with octic potential for the $\gamma$-unstable (Octic $\&$ $\gamma$-unstable) and prolate (Octic $\&$ prolate) cases, it is observed that these are very small for both approaches indicating an overall good description of the experimental data for all the isotopes considered. Moreover, these values are close for the two models making difficult to decide the most favorable description of an isotope. This latter remark leads us to preliminary assumption that these isotopes could exhibit characteristics of both types of deformations including coexistence and mixing between approximately spherical shape and $\gamma$-unstable or prolate shapes. In other words, a group of states or an entire band of an isotope can fit better the Octic $\&$ $\gamma$-unstable description than the Octic $\&$ prolate one, while other states (band) of the same isotope to prefer more an Octic $\&$ prolate structure. If this would be the case, then the complementary view on the results of the two approaches can prove to be more inspired than selecting the best description which ultimately to be the only one associated to the isotope. This method of investigation will allow us to extract more knowledge about the structure of the states corresponding to these isotopes. In the following, because there are too many calculations and data, a separate discussion is made for each isotope providing in this way a clearer view on the results.

\subsection{Discussion for $^{98}$Ru}
\begin{figure*}
	\centering
\includegraphics[width=1.0\textwidth]{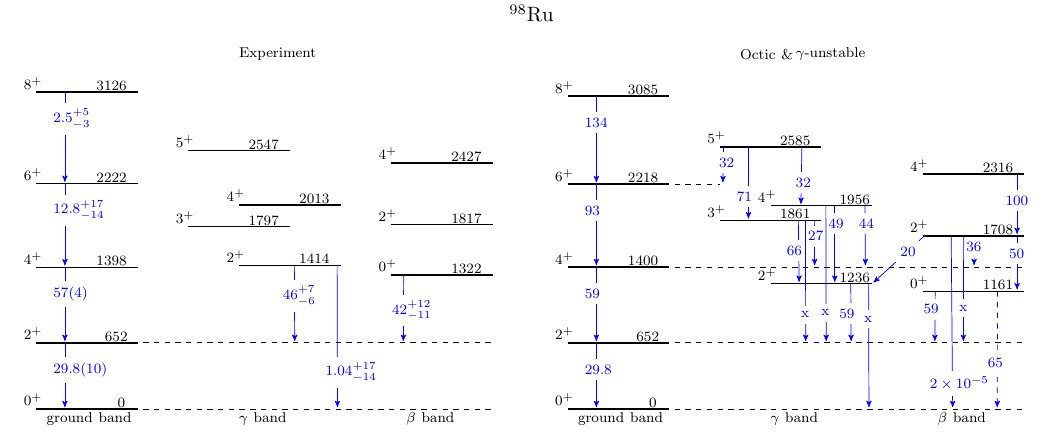}
	\caption{(Color online) The energy spectrum and the electromagnetic transitions for $^{98}$Ru calculated with the Bohr-Mottelson Hamiltonian with octic potential for $\gamma$-unstable system (Octic $\&$ $\gamma$-unstable) are compared with the corresponding experimental data (Experiment) \cite{Chen}. The energies are given in keV units, while the $B(E2)$ transitions in W.u.. The forbidden $B(E2)$s are indicated by $\mathrm{x}$, while the monopole transition between the first excited $0^+$ and the ground state by a dashed arrow. }
	\label{fig3}
\end{figure*}

According to Tables \ref{tab3} and \ref{tab4}, the $rms$ values for $^{98}$Ru are 0.148 and 0.236 for Octic $\&$ $\gamma$-unstable and Octic $\&$ prolate, respectively, tipping the balance more towards the former approach. Indeed, the data shown in Figs. \ref{fig3} and \ref{fig5} reveal, for example, a better description of the ground band energies given by the Octic $\&$ $\gamma$-unstable approach which follows closely the experimental increasing rate in energy. Instead, for Octic $\&$ prolate, the $4_{g}^{+}$ state is too high in energy, while that of the $8_{g}^{+}$ is contracting too much. The $\gamma$-unstable structure for the ground band is supported also by the experimental ratio $R_{4/2}=2.14$, which places this isotope between the spherical vibrator limit ($R_{4/2}=2.00$) and the E(5) critical point ($R_{4/2}=2.20$) of the spherical vibrator to $\gamma$-unstable phase transition.

\begin{figure*}
	\centering
\includegraphics[width=0.9\textwidth]{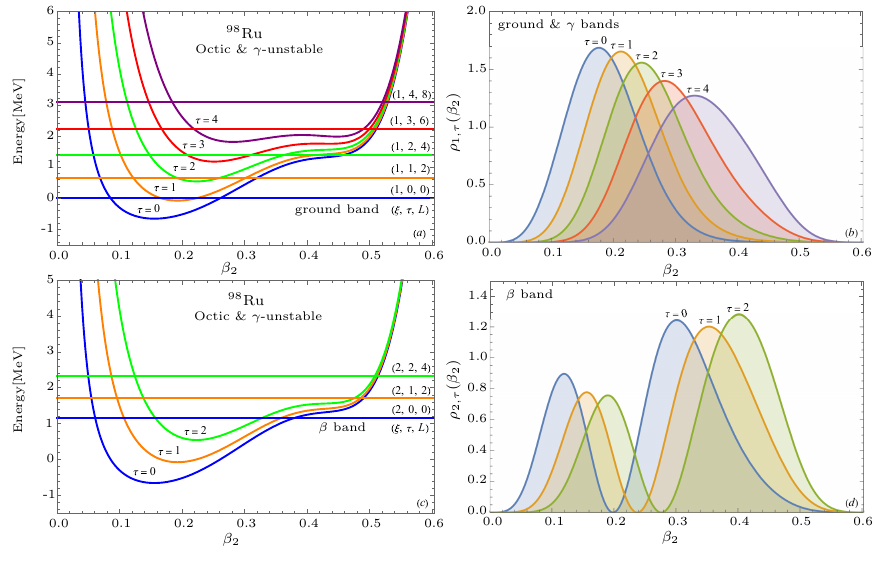}
	\caption{(Color online) The corresponding energies, effective potentials and probability density distributions of deformation for $^{98}$Ru, calculated with Octic $\&$ $\gamma$-unstable approach, are given for the states of the ground band in panels (a) and (b), respectively of the $\beta$ band in panels (c) and (d).}
	\label{fig4}
\end{figure*}

\begin{figure*}
	\centering
\includegraphics[width=1.0\textwidth]{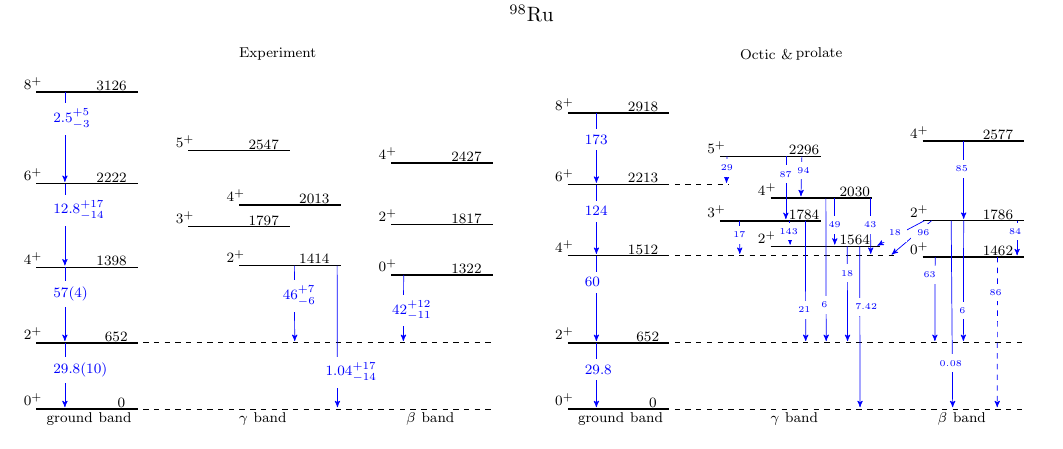}
	\caption{(Color online) The energy spectrum and the electromagnetic transitions for $^{98}$Ru calculated with the Bohr-Mottelson Hamiltonian with octic potential for $\gamma$-stable prolate system (Octic $\&$ prolate) are compared with the corresponding experimental data (Experiment) \cite{Chen}. The energies are given in keV units, while the $B(E2)$ transitions in W.u.. The monopole transition between the first excited $0^+$ and the ground state is indicated by a dashed arrow.  }
	\label{fig5}
\end{figure*}

\begin{figure*}
	\centering
\includegraphics[width=0.9\textwidth]{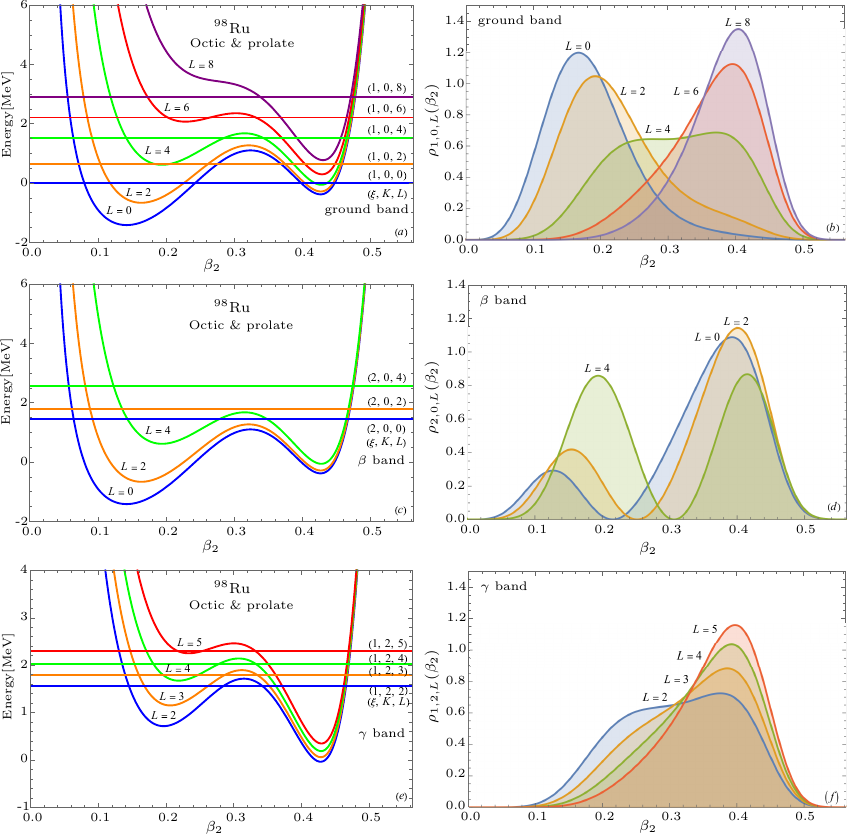}
	\caption{(Color online) The corresponding energies, effective potentials and probability density distributions of deformation for $^{98}$Ru, calculated with Octic $\&$ prolate approach, are given for the states of the ground band in panels (a) and (b), of the $\beta$ band in panels (c) and (d), and of the $\gamma$ band in panels (e) and (f), respectively. }
	\label{fig6}
\end{figure*}

Concerning the $B(E2)$ transitions in the ground band, only the first two are well reproduced with the mention that the first one is fixed for the scaling parameter $\beta_{M}$. The other two $B(E2)$s drop sharply in Experiment, while in Theory continue to grow at least at the same rate. This indicates that other phenomena than those considered are behind this behavior. Moving forward with the analysis to the $\gamma$ band energies, one remarks that the Octic $\&$ $\gamma$-unstable approach reproduces well the staggering of these states. Surprisingly, also the Octic $\&$ prolate approach is closer to this grouping of the $\gamma$ band states even if, strictly speaking, there is no staggering behavior in the $\gamma$-stable prolate case \cite{McCutchan}. If in the $\gamma$-unstable case this is the typical expected staggering, some explanations are required for that obtained with the Octic $\&$ prolate approach. In order to understand this, one has to start from the fact that within the Octic $\&$ prolate approach the deformation is not only of pure prolate type, being allowed coexistence and mixing between an approximately spherical shape and a prolate one. This generalization of the model can finally leads to different structures of the states, other than those corresponding to the limiting spherical and prolate dynamical symmetries. This behavior of the $\gamma$ band states is very clearly reflected in the plots from panels (e) and (f) of Fig. \ref{fig6}, where the probability density distribution in the $\beta$ deformation for the $2_{\gamma}^{+}$ and $3_{\gamma}^{+}$ states present forms of two peaks above the two minima of the corresponding potentials. The highest $4_{\gamma}^{+}$ and $5_{\gamma}^{+}$ states are positioned more above the well-deformed minimum, completing thus this dynamical change in shape. Another important aspect is that the experimental $B(E2;2_{\gamma}^{+}\rightarrow2_{g}^{+})=46_{-6}^{+7}$W.u. and $B(E2;0_{\beta}^{+}\rightarrow2_{g}^{+})=42_{-11}^{+12}$W.u. are very large compared to the $B(E2)$s between states of the same band being also in favor of a presence of the shape coexistence with mixing phenomena. The calculated values for the $B(E2)$ transition are 59 W.u. ($\gamma$-unstable) and 18 W.u. (prolate). The first one of 59 W.u. is very close to the experimental value $46_{-6}^{+7}$W.u. and equal with that calculated for the $B(E2;4_{g}^{+}\rightarrow2_{g}^{+})$ transition. The reason is quite obvious. The two states $4_{g}^{+}$ and $2_{\gamma}^{+}$ belong to the same $\tau=1$ multiplet within the Octic $\&$ $\gamma$-unstable approach, having thus the same set of quantum numbers. Also, the plots for energies, effective potentials and probability density distribution of deformation, shown in panels (a) and (b) of Fig. \ref{fig4}, suggest a spherical vibrator like structure for the lowest states (up to $\tau=2$) and shape fluctuations for the highest two states.  On the other hand, within the Octic $\&$ prolate case the $4_{g}^{+}$ and $2_{\gamma}^{+}$ states are no more degenerated and consequently the value of 18 W.u., which is still large and not too far from the experimental data, needs a different explanation. The answer comes quickly if one looks to the plots of the probability density distribution of deformation, given in panels (b) and (f) of Fig. \ref{fig6}, for the two involved states. For example, the plot for the $2_{g}^{+}$ state has a clear peak, but also a tail extending significantly toward the well-deformed minimum, while the plot for the $2_{\gamma}^{+}$ state has a two-peak like form even if one has $\xi=1$ (no-node) for this states. Thus, the two states present different dominant deformations given by theier highest peaks, but also a mixing between the two deformations due to the presence of a tail and of a smaller peak, respectively. This picture is associated in the frame of this approach with the presence of the shape coexistence with mixing. A similar discussion can be made also for the $B(E2;0_{\beta}^{+}\rightarrow2_{g}^{+})$ transition, of which calculated value of $59$ W.u. is more related to the $\xi=2$ vibration quantum number which produces two peaks for the probability density distribution of deformation of the $0_{\beta}^{+}$ state. This distribution in the $\beta$ deformation, shown in panel (d) of Fig. \ref{fig6},  contributes to an increased overlap for the matrix elements of the two states which is translated in an increased value for the corresponding $B(E2)$ transition. In the prolate case, this value is larger, of 63 W.u., because it is repeating the scenario from the previous transition. This time, the $0_{\beta}^{+}$ state presents a more dominant peak centered around the well-deformed minimum and a small peak above the less-deformation looking like in the mirror with the peaks of the $2_{g}^{+}$ state. This picture is not common for a typical $\beta$-vibration associated to $\xi=2$, for which the two peaks are relatively closer in height. Instead, this picture is much common with the presence of the shape coexistence with mixing. Both states presents distinct deformations, but in the same time a mixing of the two deformations \cite{BudacaA,Buganu3}.
For the smaller experimental $B(E2;2_{\gamma}^{+}\rightarrow0_{g}^{+}=1.04_{-14}^{+17}$W.u., which is forbidden (x) in the Octic $\&$ $\gamma$-unstable case, one can attribute a prolate nature since the Octic $\&$ prolate approach gives a value of 7.42 W.u.. For the $\beta$ band, the energies of the first two states $0_{\beta}^{+}$ and $2_{\beta}^{+}$ are better reproduced by Octic $\&$ prolate, while that of the $4_{\beta}^{+}$ by Octic $\gamma$-unstable.  Thus, as an intermediate conclusion, while the ground band fit better the Octic $\&$ $\gamma$-unstable configuration, the $\gamma$ and $\beta$ bands of the $^{98}$Ru present characteristics of the both configurations with shape coexistence, mixing and fluctuations signatures. This conclusion is also supported by some recent microscopic calculations \cite{Garrett1}, which found different degrees of triaxial deformation for the ground state and the head state of the $\gamma$ band, respectively a prolate shape for the head state of the $\beta$ band.
For this isotope, as well as for the next isotopes, are calculated additional electromagnetic transitions compared to the available experimental data, which can serve as reference for future experiments and other theoretical models. Also, it is calculated the monopole $E0$ transition between the first excited $0^{+}$ state and the ground state. The monopole transition is represented in figures by a dashed arrow. The experimental value for this transition is missing in the case of the $^{98}$Ru. The calculated ones, $65$ ($\gamma$-unstable) and $86$ (prolate), indicate a possible presence of shape mixing \cite{Wood}.

\subsection{Discussion for $^{100}$Ru}
\begin{figure*}
	\centering
\includegraphics[width=1.0\textwidth]{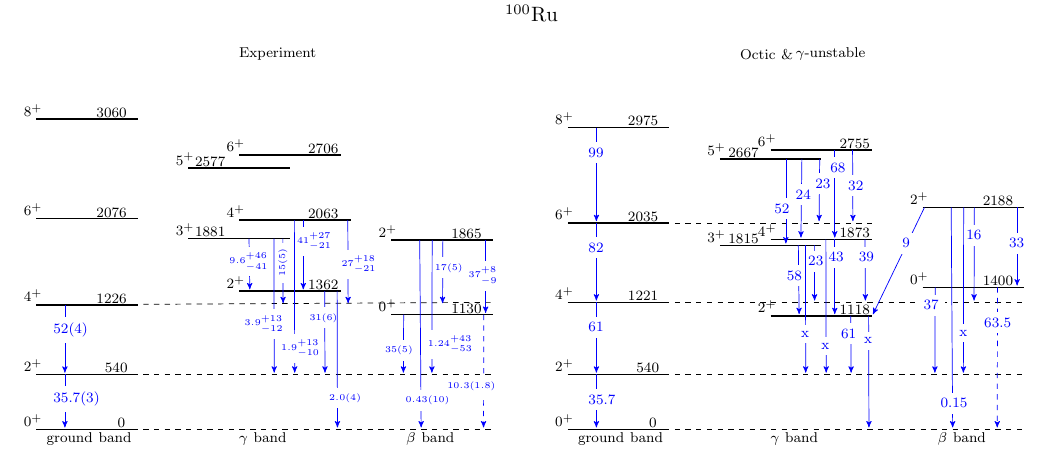}
	\caption{(Color online) The energy spectrum and the electromagnetic transitions for $^{100}$Ru calculated with the Bohr-Mottelson Hamiltonian with octic potential for $\gamma$-unstable system (Octic $\&$ $\gamma$-unstable) are compared with the corresponding experimental data (Experiment) \cite{Singh}. The energies are given in keV units, while the $B(E2)$ transitions in W.u.. The forbidden $B(E2)$s are indicated by $\mathrm{x}$, while the monopole transition between the first excited $0^+$ and the ground state by a dashed arrow. }
	\label{fig7}
\end{figure*}

The $rms$ values for the $^{100}$Ru isotope, given in Tables \ref{tab3} and \ref{tab4}, are 0.315 ($\gamma$-unstable) and 0.190 (prolate). Thus, one can expect to see more prolate fingerprints for this isotope than $\gamma$-unstable ones. According to Figs. \ref{fig7} and \ref{fig9}, this time the agreement with the energies of the ground band is very good for both approaches with a slightly better description offered by the Octic $\&$ $\gamma$-unstable approach for the $4_{g}^{+}$ and $6_{g}^{+}$ states. The experimental difference in energy between the first two states of the $\beta$ band is of 735 keV, being estimated to be of 788 keV ($\gamma$-unstable) and 785 keV (prolate). Thus, both theoretical values are close to the experimental value with the mention that the Octic $\&$ prolate approach predicts well also the scaling of the two energies of the $\beta$ band. Concerning the energies of the $\gamma$-band it is clear that the Octic $\&$ $\gamma$-unstable approach reproduces well the staggering (grouping) of the states, while the Octic $\&$ prolate approach  doesn't keep the step with this staggering.  Based only on the analysis made for the energy levels, one can say that the ground band is almost equally disputed between the two approaches, while the balance is in the favor of a more prolate structure for the $\beta$ band, respectively of a more $\gamma$-unstable structure for the $\gamma$ band.

\begin{figure*}
	\centering
\includegraphics[width=0.9\textwidth]{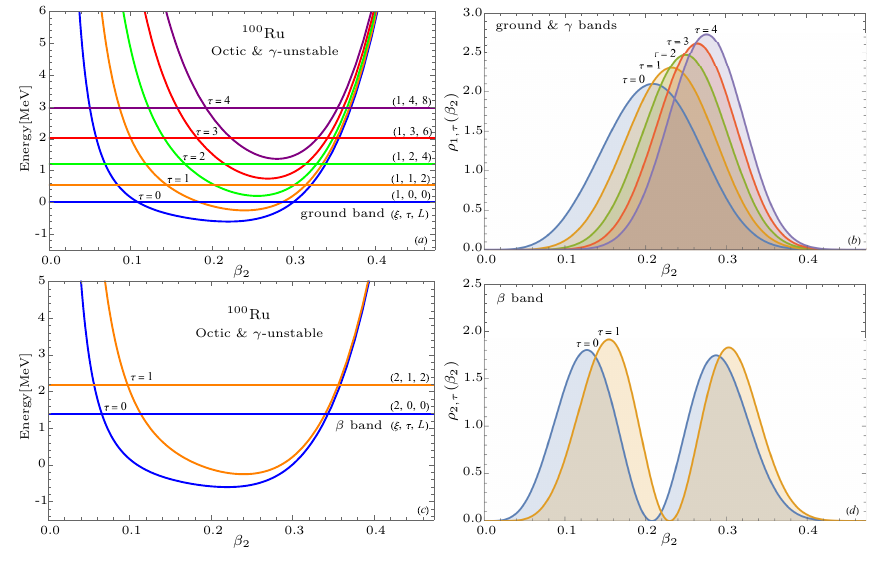}
	\caption{(Color online) The corresponding energies, effective potentials and probability density distributions of deformation for $^{100}$Ru, calculated with Octic $\&$ $\gamma$-unstable approach, are given for the states of the ground band in panels (a) and (b), respectively of the $\beta$ band in panels (c) and (d). }
	\label{fig8}
\end{figure*}

\begin{figure*}
	\centering
\includegraphics[width=1.0\textwidth]{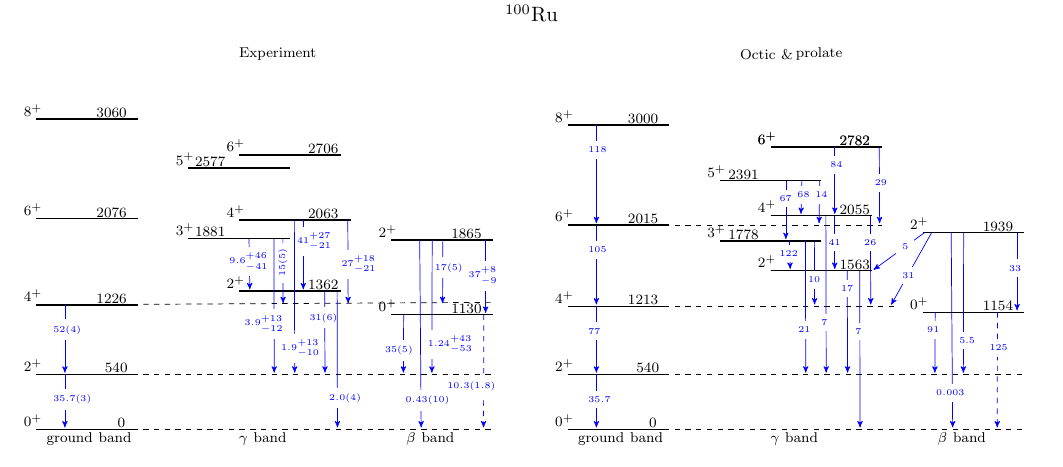}
	\caption{(Color online) The energy spectrum and the electromagnetic transitions for $^{100}$Ru calculated with the Bohr-Mottelson Hamiltonian with octic potential for $\gamma$-stable prolate system (Octic $\&$ prolate) are compared with the corresponding experimental data (Experiment) \cite{Singh}. The energies are given in keV units, while the $B(E2)$ transitions in W.u.. The monopole transition between the first excited $0^+$ and the ground state is indicated by a dashed arrow. }
	\label{fig9}
\end{figure*}

\begin{figure*}
	\centering
\includegraphics[width=0.9\textwidth]{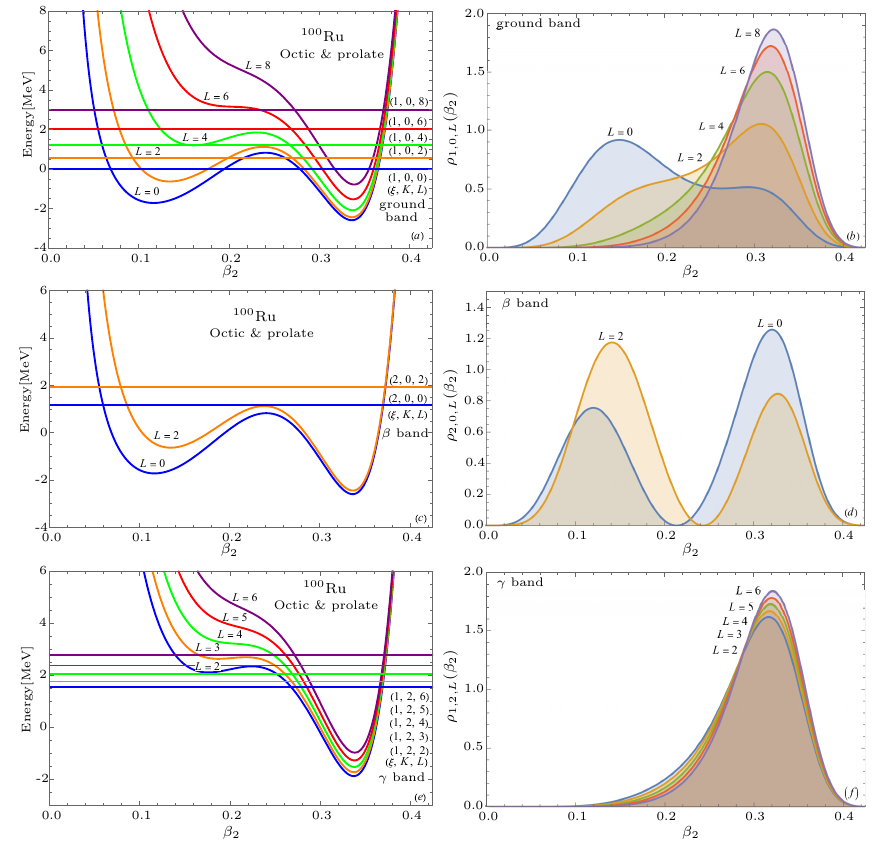}
	\caption{(Color online) The corresponding energies, effective potentials and probability density distributions of deformation for $^{100}$Ru, calculated with Octic $\&$ prolate approach, are given for the states of the ground band in panels (a) and (b), of the $\beta$ band in panels (c) and (d), and of the $\gamma$ band in panels (e) and (f), respectively.}
	\label{fig10}
\end{figure*}

A further investigation of the electromagnetic transitions in relation to other quantities, such as the effective potentials and the probability distribution of the $\beta$ deformation shown in Figs. \ref{fig8} and \ref{fig10}, is expected to make more light on the structure of these states. In the ground band there are only two $B(E2)$ transitions, the first one $B(E2;2_{g}^{+}\rightarrow0_{g}^{+})=35.7(3)$ W.u. is fixed such that to determine the scaling parameter $\beta_{M}$ of the transition operator. Thus, the only one which is effectively calculated for this band is the $B(E2;4_{g}^{+}\rightarrow2_{g}^{+})$ transition getting 61 W.u. ($\gamma$-unstable) and 77 W.u. (prolate). The former value is much closer to the experimental value of $52(4)$ W.u., especially if one takes into account the upper measurement error. Moreover, the experimental energy ratio $R_{4/2}=2.27$ gives this isotope as good candidate for the critical point of the spherical vibrator to $\gamma$-unstable phase transition. This quantity is calculated to be 2.26 ($\gamma$-unstable) and 2.25 (prolate), respectively. Thus, finally the ground band seems to fit better a $\gamma$-unstable configuration being closer to the E(5) critical point picture. This is supported also by the flat shape of the effective potentials, respectively the more extended peaks of the lowest states, shown in panels (a) and (b) of Fig. \ref{fig8}. The $B(E2)$s in the $\beta$ band and from the $\beta$ band to the ground band are very well reproduced, especially within the Octic $\&$ $\gamma$-unstable approach. Excepting the $B(E2;2_{\beta}^{+}\rightarrow2_{g}^{+})$ transition, which is small in experiment but not enough to be considered forbidden and which can be explained by a prolate structure, the other $B(E2)$s clearly indicate here a $\gamma$-unstable configuration. Also, the monopole transition of 10.3(1.8) is much better reproduced by the Octic $\&$ $\gamma$-unstable approach (63.5) than the Octic $\&$ prolate approach (125). This large value predicted for the monopole transition within the Octic $\&$ prolate, as well as for the $B(E2;0_{\beta}^{+}\rightarrow2_{g}^{+})$ of 91 W.u. are related to the presence of two peaks for the states $0_{g}^{+}$ and $2_{g}^{+}$, shown in panel (a) of Fig. \ref{fig10}, which represents a signature for the shape coexistence with mixing. On the other hand, it is clear that these values are not so close to the corresponding experimental data compared to those calculated with the Octic $\&$ $\gamma$-unstable approach. Thus, the appearance of these two peaks for $0_{g}^{+}$ and $2_{g}^{+}$, respectively with plots of the probability density distribution of deformation for the $0_{\beta}^{+}$ and $2_{\beta}^{+}$ different than those for a regular $\beta$-prolate vibration, is related to the more precise prediction of the $\beta$ band energies by the Octic $\&$ prolate approach. Indeed, it is well known that a low energy for the first excited $0^{+}$ state, $\triangle E=E_{0_{2}^{+}}-E_{2_{g}^{+}}<800$ keV (590 kev for $^{100}$Ru), is associated with the presence of the shape coexistence \cite{Martinou2}. In the Octic $\&$ $\gamma$-unstable case, the energies of the $\beta$ band are shifted up, not too much but enough so that the coexistence goes unnoticed. This is not the case within the Octic $\&$ prolate approach, where the energy of the first excited $0^{+}$ is very well reproduced with a deviation of about $2\%$. Therefore, it would be more appropriate to admit for the $\beta$ band a mixing of the $\gamma$-unstable and prolate configurations, respectively including a coexistence with mixing between an approximately spherical shape and a more deformed one. A very good agreement is also obtained for the $B(E2)$s between the states of the $\gamma$ band, respectively between the $\gamma$ band and the ground band. Comparing the results of the two approaches, excepting for the $B(E2;3_{\gamma}^{+}\rightarrow2_{\gamma}^{+})$, the Octic $\&$ prolate approach is very close to the experimental data, and moreover gives appropriate values for those $B(E2)$s which are forbidden within the Octic $\&$ $\gamma$-unstable approach. The probability density distribution of deformation for the states of the $\gamma$ band presented in panel (b) of Fig. \ref{fig8}, together with the effective potentials from panel (a) of the same figure, show a $\gamma$-unstable picture, which is in agreement with the experimental staggering of the states. On the other hand, in panels (e) and (f) of Fig. \ref{fig10}, the plots for these quantities show a deeply deformed minimum preferred by the states of the $\gamma$ band, all of them presenting a single peak above this more deformed minimum. On the contrary, for the ground state one has two peaks, in panel (b) of Fig. (\ref{fig10}), with the higher peak above the less deformed minimum. The $2_{g}^{+}$ makes the shape transition toward the well-deformed minimum, having also two peaks, but with the highest peak above the well-deformed minimum. The states $4_{g}^{+}$, $6_{g}^{+}$ and $8_{g}^{+}$ complete the dynamical shape transition, all of them having a single peak centered more above the well-deformed minimum. Concerning the plots of the probability density distribution of deformation for the two states of the $\beta$ band, given in panel (d) of Fig. \ref{fig10}, these show an opposite vibration of one state compared to the other state. These distributions in $\beta_2$ deformation of the states correspond to the presence of the shape coexistence and mixing. In conclusion, a complementary application of the two approaches leads to a better interpretation and understanding of the corresponding experimental data of the $^{100}$Ru, emerging the presence of the shape coexistence and mixing phenomena.

\subsection{Discussion for $^{102}$Ru}
\begin{figure*}
	\centering
\includegraphics[width=1.0\textwidth]{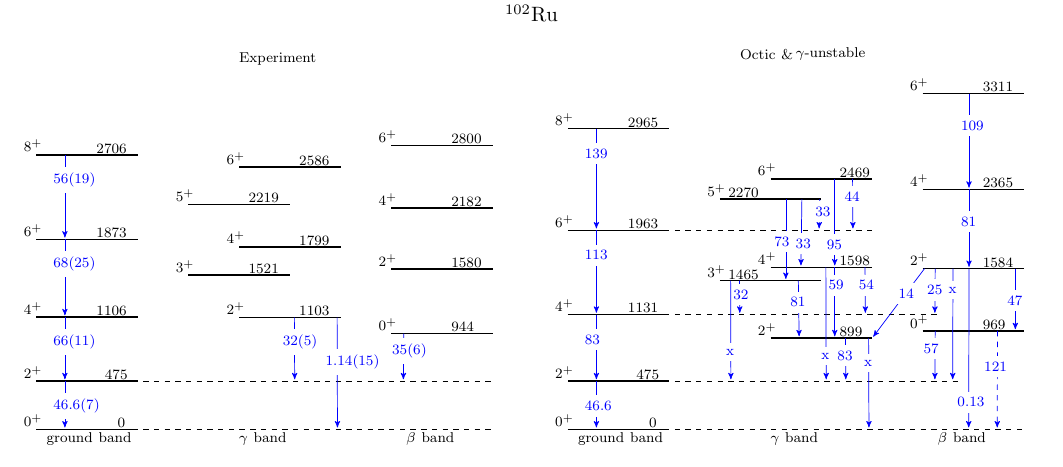}
	\caption{(Color online) The energy spectrum and the electromagnetic transitions for $^{102}$Ru calculated with the Bohr-Mottelson Hamiltonian with octic potential for $\gamma$-unstable system (Octic $\&$ $\gamma$-unstable) are compared with the corresponding experimental data (Experiment) \cite{Frenne1}. The energies are given in keV units, while the $B(E2)$ transitions in W.u.. The forbidden $B(E2)$s are indicated by $\mathrm{x}$, while the monopole transition between the first excited $0^+$ and the ground state by a dashed arrow. }
	\label{fig11}
\end{figure*}

\begin{figure*}
	\centering
\includegraphics[width=0.9\textwidth]{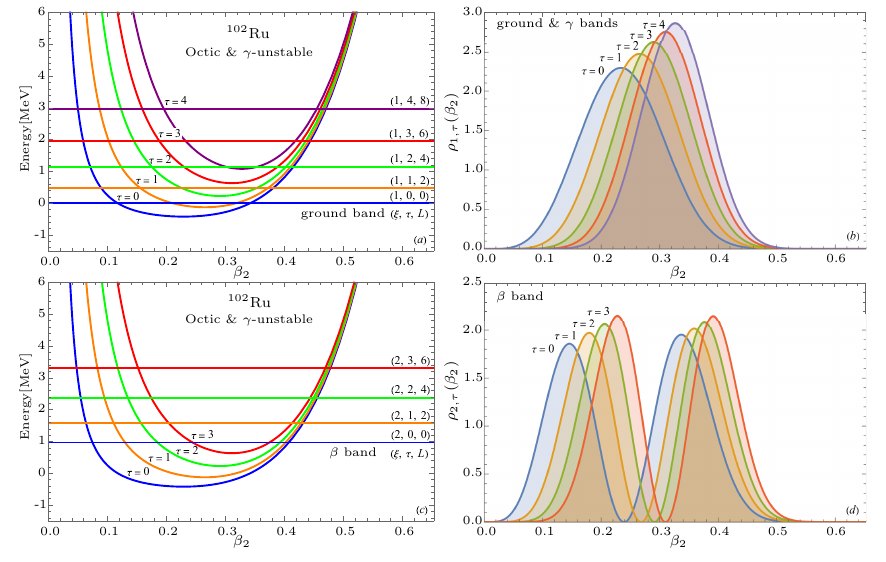}
	\caption{(Color online) The corresponding energies, effective potentials and probability density distributions of deformation for $^{102}$Ru, calculated with Octic $\&$ $\gamma$-unstable approach, are given for the states of the ground band in panels (a) and (b), respectively of the $\beta$ band in panels (c) and (d). }
	\label{fig12}
\end{figure*}

\begin{figure*}
	\centering
\includegraphics[width=1.0\textwidth]{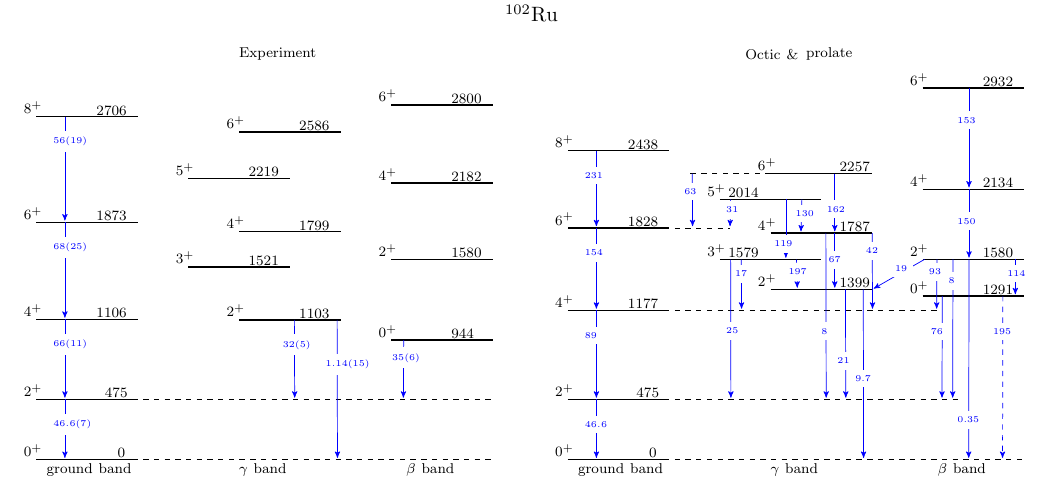}
	\caption{(Color online) The energy spectrum and the electromagnetic transitions for $^{102}$Ru calculated with the Bohr-Mottelson Hamiltonian with octic potential for $\gamma$-stable prolate system (Octic $\&$ prolate) are compared with the corresponding experimental data (Experiment) \cite{Frenne1}. The energies are given in keV units, while the $B(E2)$ transitions in W.u.. The monopole transition between the first excited $0^+$ and the ground state is indicated by a dashed arrow. }
	\label{fig13}
\end{figure*}

Most of the energies of this isotope, presented in Figs. \ref{fig11} and \ref{fig13}, seem to be better described by the Octic $\&$ $\gamma$-unstable approach. Moreover, the staggering of the $\gamma$ band is reproduced only by the Octic $\&$ $\gamma$-unstable approach, as well as the energy of the $0_{\beta}^{+}$ state. The energy of the $2_{\beta}^{+}$ state is exactly determined by the Octic $\&$ prolate approach, respectively with a very small deviation of only $4$ keV by the Octic $\&$ $\gamma$-unstable approach, while the energies of the other two states, $4_{\beta}^{+}$ and $6_{\beta}^{+}$, fit better the Octic $\&$ prolate description. Thus, one can consider more a $\gamma$-unstable configuration for this isotope with a possible presence of a prolate structure for the $2_{\beta}^+$, $4_{\beta}^+$ and $6_{\beta}^+$ states. The effective potentials of a flat shape, as well as plots of the probability density distribution of the $\beta_{2}$ deformation, shown in Fig. \ref{fig12}, recommend this isotope as a good candidate for the critical point of the spherical vibrator to $\gamma$-unstable phase transition. This is in agreement with the results obtained in \cite{Lahbas} for this isotope using a quasi-exactly solvable method for the Bohr-Mottelson Hamiltonian with sextic oscillator potential. Concerning the $B(E2)$s, the experimental data manifest a contraction with the increasing of the spin, which is not seen in theory. Both approaches indicate an increasing of these values, with a higher rate within the Octic $\&$ prolate approach. The $B(E2;0_{\beta}^{+}\rightarrow2_{g}^{+})=35(6)$ W.u. is calculated at 57 W.u. ($\gamma$-unstable) and 76 W.u. (prolate), being closer to the picture of the $\gamma$-unstable structure. On the other hand, for the only two available experimental $B(E2)$s from the $\gamma$ band to the ground band, the Octic $\&$ prolate approach is more appropriate with the value of 21 W.u. for $B(E2;2_{\gamma}^{+}\rightarrow2_{g}^{+})=32(5)$ W.u. than of 83 W.u. ($\gamma$-unstable), respectively with 9.7 W.u. for $B(E2;2_{\gamma}^{+}\rightarrow0_{g}^{+})=1.14(15)$ W.u. than forbidden  within the Octic $\&$ $\gamma$-unstable approach due to the selection rules given by Eq. (\ref{tauop}). The value of the $B(E2;2_{\gamma}^{+}\rightarrow2_{g}^{+})$ transition is higher and equal to the $B(E2;4_{g}^{+}\rightarrow2_{g}^{+})$ within the Octic $\&$ $\gamma$-unstable approach because the $4_{g}^{+}$ and $2_{\gamma}^{+}$ are degenerated both belonging to the same multiplet given by $\tau=2$. This is not the case within the Octic $\&$ prolate approach.

\begin{figure*}
	\centering
\includegraphics[width=0.9\textwidth]{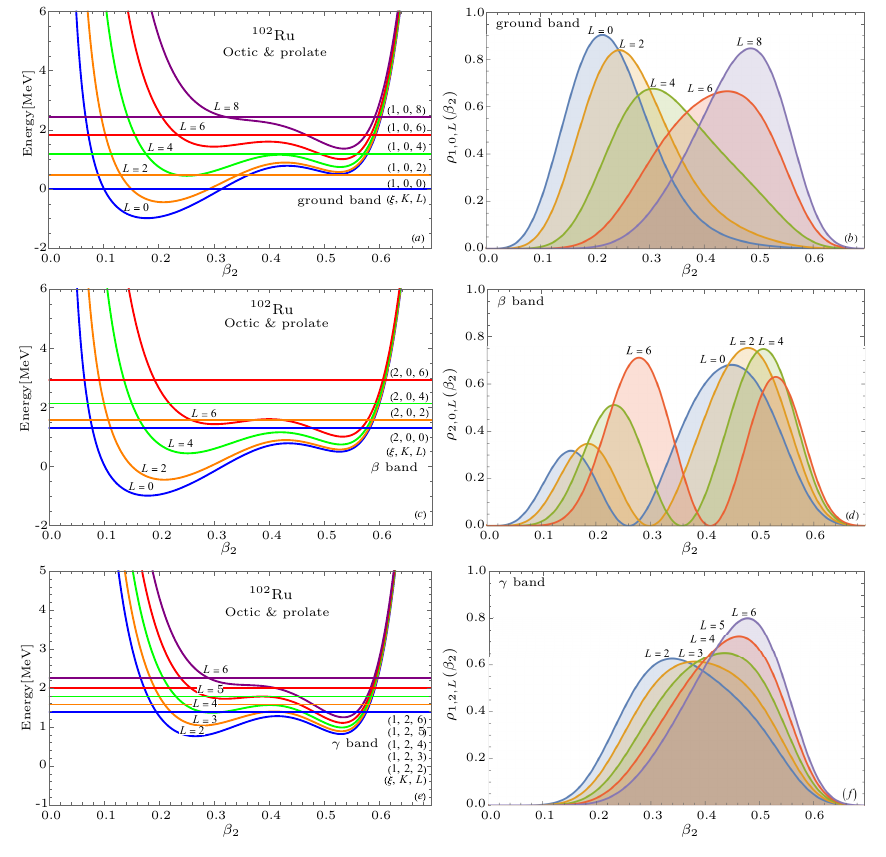}
	\caption{(Color online) The corresponding energies, effective potentials and probability density distributions of deformation for $^{102}$Ru, calculated with Octic $\&$ prolate approach, are given for the states of the ground band in panels (a) and (b), of the $\beta$ band in panels (c) and (d), and of the $\gamma$ band in panels (e) and (f), respectively.}
	\label{fig14}
\end{figure*}
Moreover, within the Octic $\&$ prolate approach, the effective potential for the $2_{\gamma}^{+}$, from panel (e) of Fig. \ref{fig14}, has two almost degenerated minima separated by a small barrier (maximum). Also, the plot for the probability density distribution of deformation of the $2_{\gamma}^{+}$, from panel (f), has an extended peak over the two minima of the potential. This picture corresponds in the present model to the shape fluctuation case \cite{BudacaA}. The corresponding peak for the $2_{g}^{+}$, from panel (b) of Fig. \ref{fig14}, is more centered around the less deformed minimum, but with a large width and a tail extension above the well deformed minimum. These distributions of the deformation for the $2_{\gamma}^{+}$ and $2_{g}^{+}$ states lead to a consistent overlap between them, respectively to a good mixing due to the shape fluctuations, and finally to this closely predicted value to the corresponding experimental data. The peak for the $0_{g}^{+}$ is more on the left side than the one for $2_{g}^{+}$ in relation to that for $2_{\gamma}^{+}$, resulting thus in a smaller overlap, respectively to a smaller $B(E2)$ value. The shape fluctuations emerge also for the $3_{\gamma}^{+}$ and $4_{\gamma}^{+}$ states which still preserve the two minima close in energy, respectively extended peaks for the probability density distribution of deformation (panels (e) and (f) of Fig, \ref{fig14}). For higher states of the $\gamma$ band, the first minimum vanishes and the states start to have a more localized deformation than to manifest a shape fluctuation.

\subsection{Discussion for $^{104}$Ru}
\begin{figure*}
	\centering
\includegraphics[width=1.0\textwidth]{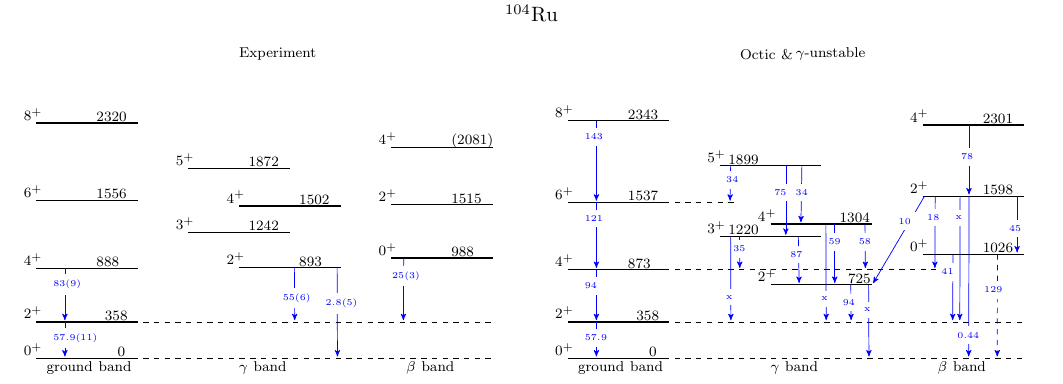}
	\caption{(Color online) The energy spectrum and the electromagnetic transitions for $^{104}$Ru calculated with the Bohr-Mottelson Hamiltonian with octic potential for $\gamma$-unstable system (Octic $\&$ $\gamma$-unstable) are compared with the corresponding experimental data (Experiment) \cite{Blachot}. The energies are given in keV units, while the $B(E2)$ transitions in W.u.. The forbidden $B(E2)$s are indicated by $\mathrm{x}$, while the monopole transition between the first excited $0^+$ and the ground state by a dashed arrow. }
	\label{fig15}
\end{figure*}

\begin{figure*}
	\centering
\includegraphics[width=0.9\textwidth]{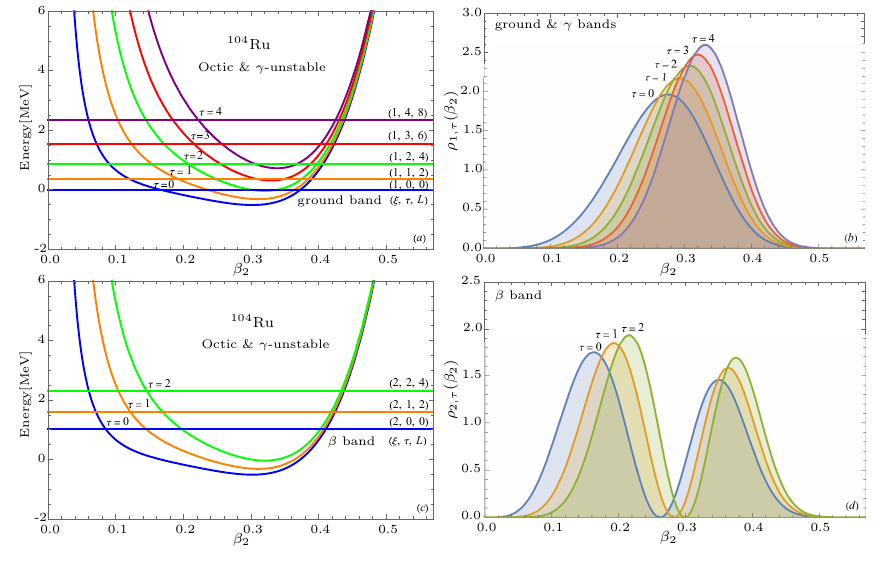}
	\caption{(Color online) The corresponding energies, effective potentials and probability density distributions of deformation for $^{104}$Ru, calculated with Octic $\&$ $\gamma$-unstable approach, are given for the states of the ground band in panels (a) and (b), respectively of the $\beta$ band in panels (c) and (d). }
	\label{fig16}
\end{figure*}

\begin{figure*}
	\centering
\includegraphics[width=1.0\textwidth]{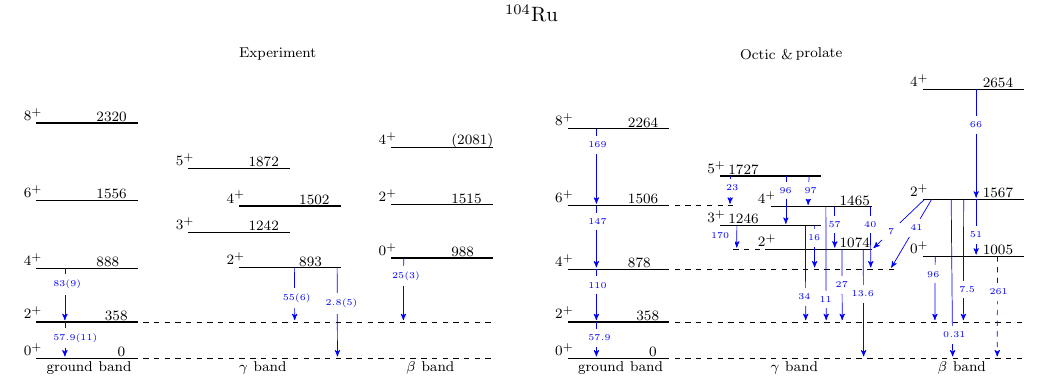}
	\caption{(Color online) The energy spectrum and the electromagnetic transitions for $^{104}$Ru calculated with the Bohr-Mottelson Hamiltonian with octic potential for $\gamma$-stable prolate system (Octic $\&$ prolate) are compared with the corresponding experimental data (Experiment) \cite{Blachot}. The energies are given in keV units, while the $B(E2)$ transitions in W.u.. The monopole transition between the first excited $0^+$ and the ground state is indicated by a dashed arrow.}
	\label{fig17}
\end{figure*}

\begin{figure*}
	\centering
\includegraphics[width=0.9\textwidth]{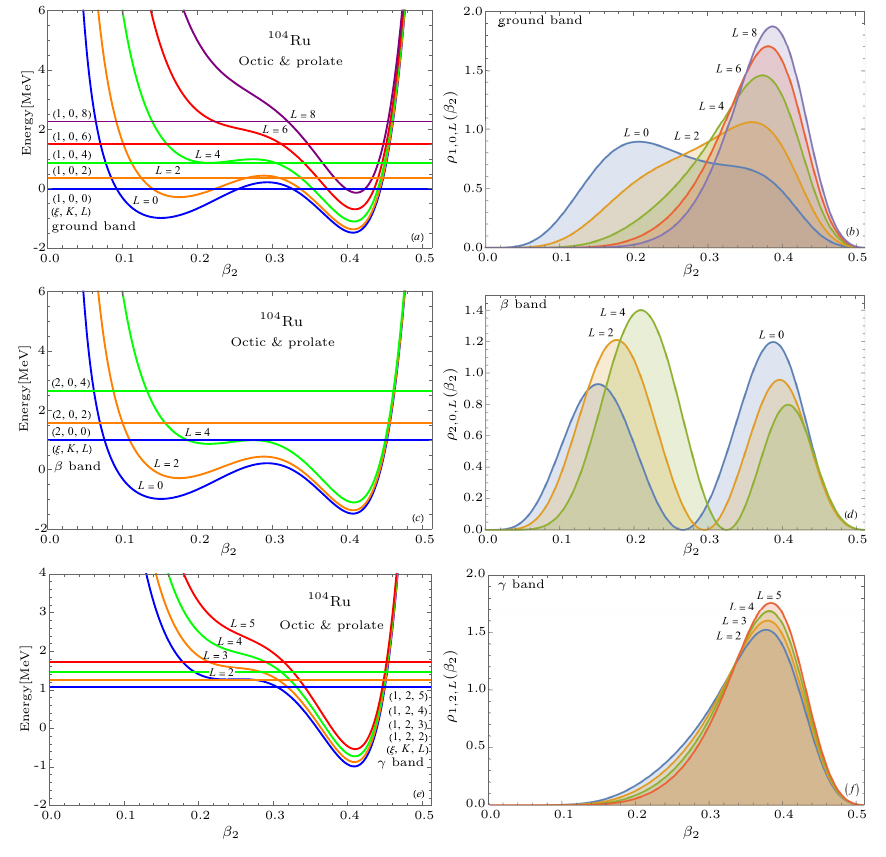}
	\caption{(Color online) The corresponding energies, effective potentials and probability density distributions of deformation for $^{104}$Ru, calculated with Octic $\&$ prolate approach, are given for the states of the ground band in panels (a) and (b), of the $\beta$ band in panels (c) and (d), and of the $\gamma$ band in panels (e) and (f), respectively.}
	\label{fig18}
\end{figure*}

For this isotope, both approaches describe well the energies of the ground and $\beta$ bands from Figs. \ref{fig15} and \ref{fig17}  with a favorable fit of the ground band for Octic $\&$ $\gamma$-unstable, respectively of the $\beta$ band for Octic $\&$ prolate. An exception is the energy of the $4_{\beta}^{+}$, which is written in brackets, being not included in the fit of the model parameters. The staggering of the $\gamma$ band states is well reproduced within the Octic $\&$ $\gamma$-unstable approach, while the Octic $\&$ prolate approach follows closely this grouping of the states but without strictly reproducing it. The two experimental $B(E2)$s, $83(9)$ W.u. and $25(3)$ W.u., between $4_{g}^{+}\rightarrow2_{g}^{+}$ and $0_{\beta}^{+}\rightarrow2_{g}^{+}$ fit better the values $94$ W.u. and $41$ W.u. of the Octic $\&$ $\gamma$-unstable approach than those of the Octic $\&$ prolate approach of $110$ W.u. and $96$ W.u.. On the other hand, the $B(E2;2_{\gamma}^{+}\rightarrow2_{g}^{+})=55(6)$ W.u. and $B(E2;2_{\gamma}^{+}\rightarrow0_{g}^{+})=2.8(5)$ W.u. transitions are better reproduced within the Octic $\&$ prolate approach with $27$ W.u., respectively 13.6 W.u.. The latter value, even if it is slightly greater than the experimental value, confirms that this transition is not experimentally forbidden. Thus, the states of the $^{104}$Ru isotope share features of both $\gamma$-unstable and prolate symmetries, non of the three bands being fully described by just one of the two involved approaches. Instead, the two approaches complement each other very well. According to Fig. \ref{fig16}, within the Octic $\&$ $\gamma$-unstable approach this isotope is on the deformed side of the E(5) critical point showing a tendency of a single deformed minimum in panels (a)-(d) of this figure. Instead, two minima appear for the potentials of the ground and $\beta$ bands within the Octic $\&$ prolate approach, in panels (a) and (c) of Fig. \ref{fig18}, with the well-deformed one lower in energy, while the less-deformed minimum almost vanishing for the $\gamma$-unstable case. These potentials forms induce a coexistence with mixing in the ground and $\beta$ bands, respectively a dynamical shape evolution in band. The ground state, in panel (b) of Fig. \ref{fig18}, has two peaks with the dominant one above the less-deformed minimum, the $2_{g}^{+}$ makes the transition between the two deformations manifesting a two-peak like distribution in deformation with the highest peak above the well-deformed minimum, while the states $4_{g}^{+}$, $6_{g}^{+}$ and $8_{g}^{+}$ have a single peak above the well-deformed minimum. This behavior of the states in relation to their deformation is called here a dynamical shape transition in band. A similar picture is observed for the states of the $\beta$ band in panel (d), this time the dynamical shape transition taking place from the well-deformed minimum $(0_{\beta}^{+})$ to the less-deformation for the next excited $2_{\beta}^{+}$ and $4_{\beta}^{+}$ states. This phenomenon is different than that associated with the ground-state shape phase transition where the change in shape takes place in relation to the ground state only from one nucleus to another.

\subsection{Discussion for $^{106}$Ru}
\begin{figure*}
	\centering
\includegraphics[width=1.0\textwidth]{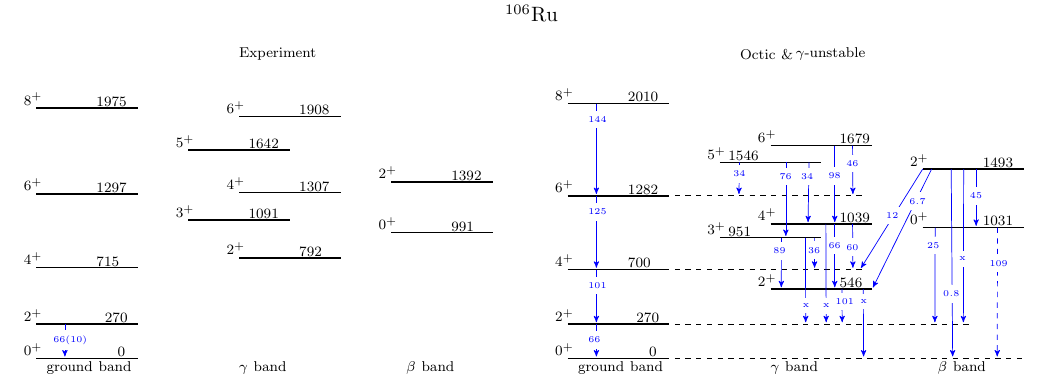}
	\caption{(Color online) The energy spectrum and the electromagnetic transitions for $^{106}$Ru calculated with the Bohr-Mottelson Hamiltonian with octic potential for $\gamma$-unstable system (Octic $\&$ $\gamma$-unstable) are compared with the corresponding experimental data (Experiment) \cite{Frenne2}. The energies are given in keV units, while the $B(E2)$ transitions in W.u.. The forbidden $B(E2)$s are indicated by $\mathrm{x}$, while the monopole transition between the first excited $0^+$ and the ground state by a dashed arrow. }
	\label{fig19}
\end{figure*}

\begin{figure*}
	\centering
\includegraphics[width=0.9\textwidth]{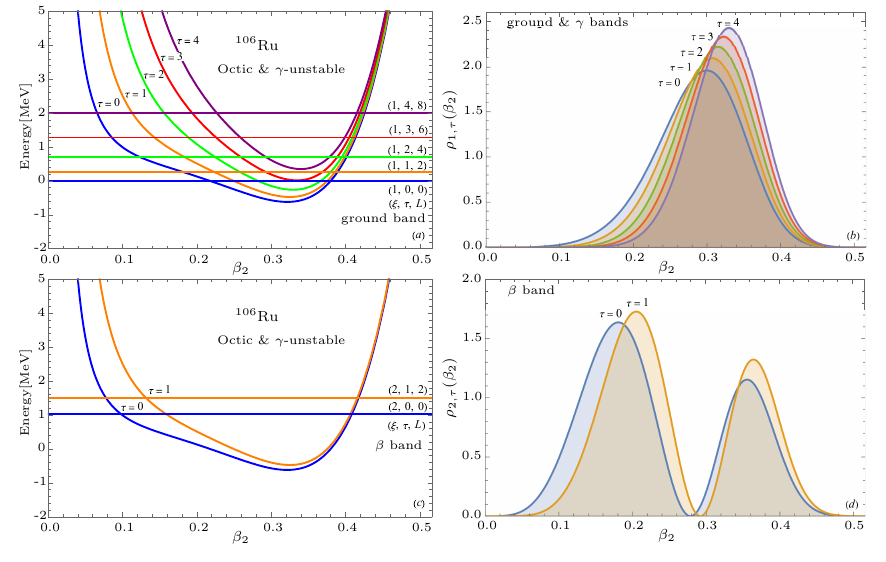}
	\caption{(Color online) The corresponding energies, effective potentials and probability density distributions of deformation for $^{106}$Ru, calculated with Octic $\&$ $\gamma$-unstable approach, are given for the states of the ground band in panels (a) and (b), respectively of the $\beta$ band in panels (c) and (d). }
	\label{fig20}
\end{figure*}

\begin{figure*}
	\centering
\includegraphics[width=1.0\textwidth]{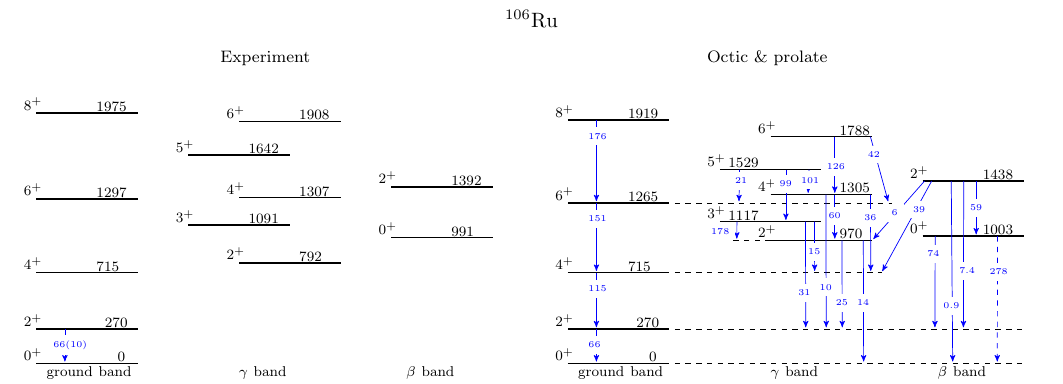}
	\caption{(Color online) The energy spectrum and the electromagnetic transitions for $^{106}$Ru calculated with the Bohr-Mottelson Hamiltonian with octic potential for $\gamma$-stable prolate system (Octic $\&$ prolate) are compared with the corresponding experimental data (Experiment) \cite{Frenne2}. The energies are given in keV units, while the $B(E2)$ transitions in W.u.. The monopole transition between the first excited $0^+$ and the ground state is indicated by a dashed arrow. }
	\label{fig21}
\end{figure*}

\begin{figure*}
	\centering
\includegraphics[width=0.9\textwidth]{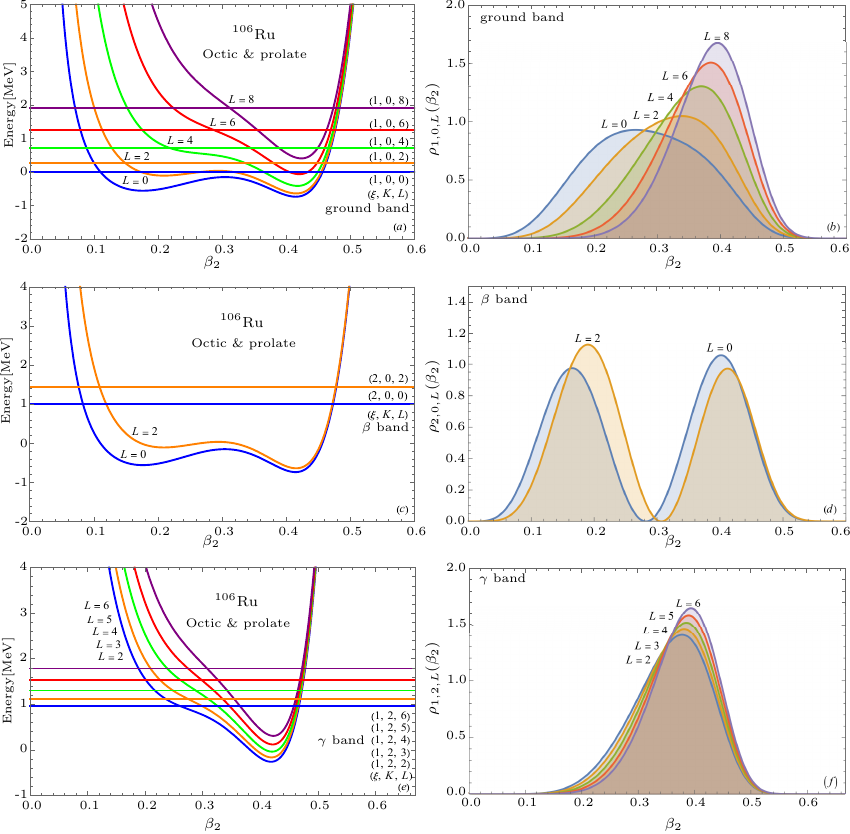}
	\caption{(Color online) The corresponding energies, effective potentials and probability density distributions of deformation for $^{106}$Ru, calculated with Octic $\&$ prolate approach, are given for the states of the ground band in panels (a) and (b), of the $\beta$ band in panels (c) and (d), and of the $\gamma$ band in panels (e) and (f), respectively. }
	\label{fig22}
\end{figure*}

\begin{figure}
	\centering
\includegraphics[width=0.45\textwidth]{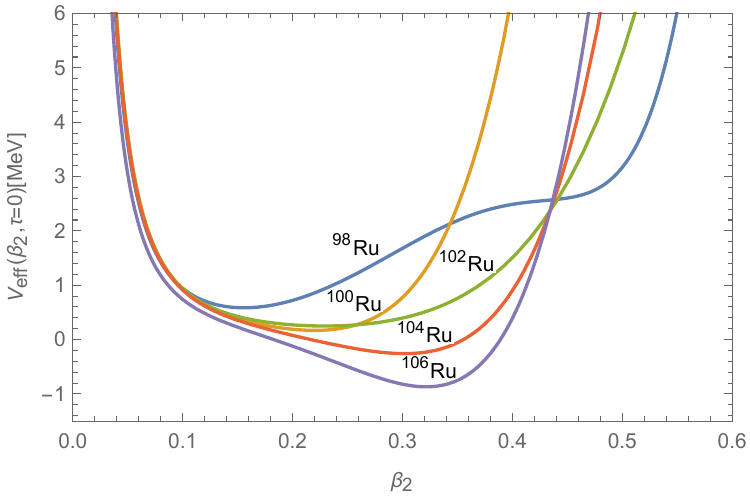}
	\caption{(Color online) The effective potentials, given by Eq. (\ref{effpot}) and $\tau=0$, within the Octic $\&$ $\gamma$-unstable approach plotted for the $^{98-106}$Ru isotopes.}
	\label{fig23}
\end{figure}

The values of the $rms$ for $^{106}$Ru are $0.558$ ($\gamma$-unstable) and 0.301 (prolate), which is reflected in the fact that the Octic $\&$ prolate approach describes better the energies shown in Figs. \ref{fig19} and \ref{fig21} of the following states: $4_{g}^{+}$, $0_{\beta}$, $2_{\beta}^{+}$, $2_{\gamma}^{+}$, $3_{\gamma}^{+}$, $4_{\gamma}^{+}$, $6_{\gamma}^{+}$. Thus, one can say that the $\beta$ band is more prolate, but not also the ground band where the energies of the highest $6_{g}^{+}$ and $8_{g}^{+}$ states fit better the Octic $\&$ $\gamma$-unstable description. Moreover, even if numerically the energies of the $\gamma$ band are closer to the experimental ones than those of the Octic $\&$ prolate approach, the staggering of the $\gamma$ band is described only by the Octic $\gamma$-unstable approach. Also, taking into account the experimental ratio $R_{4/2}=2.6$ and the fact that the effective potentials, given in panels (a) and (c) of Fig. \ref{fig20}, present a single deformed minimum, one can conclude that the ground and the $\gamma$ bands of this isotope falls into the $\gamma$-unstable limiting symmetry. This is not a coincidence and it makes sense if one looks back to the effective potentials for $\tau=0$ in the Octic $\&$ $\gamma$-unstable case. These potentials, plotted in Fig. \ref{fig23}, describe a clear phase transition from an approximately spherical vibrator ($^{98}$Ru) to a $\gamma$-unstable deformation ($^{104,106}$Ru) with the critical point for the flat potentials ($^{100,102}$Ru). This is not the case for the Octic $\&$ prolate approach, where the potentials have two minima and most often the probability density distribution of deformation in $\beta$ indicates a presence of the shape coexistence and mixing in these isotopes. This latter result is in agreement with the signature $\triangle E=E_{0_{2}^{+}}-E_{2_{g}^{+}}<800$ keV \cite{Martinou2} which is satisfied by all the considered $^{98-106}$Ru isotopes: 670 keV ($^{98}$Ru), 590 keV ($^{100}$Ru), 469 keV ($^{102}$Ru), 630 keV ($^{104}$Ru), 721 keV ($^{106}$Ru). Thus, the greatest value of this quantity corresponds to the $^{106}$Ru isotope, which is reflected also in the plots for potentials and probability density distribution of deformation shown in Fig. \ref{fig22}. For example, the potential for the ground state, from panel (a), has two minima separated by a small barrier with the ground state slightly above this barrier. This produces a single extended peak for the ground state, in panel (b), which is associated more with the shape fluctuations than with the shape coexistence.

\section{Conclusions}
The Bohr-Mottelson Hamiltonian with an octic potential in the $\beta$ variable is numerically solved in a basis of Bessel functions of the first kind for axially-symmetric deformed nuclei. This approach, conventionally called Octic $\&$ prolate, is able to describe a phase transition between the spherical vibrator and the prolate rotor, as well as shape coexistence and mixing between these two limiting symmetry configurations. The Octic $\&$ prolate approach is applied, in the present study, together with the previously proposed solution of the Bohr-Mottelson Hamiltonian with octic potential for $\gamma$-unstable symmetry (Octic $\&$ $\gamma$-unstable) to investigate the low-lying quadrupole collective states of the $^{98-106}$Ru even-even isotopes. Additional calculations are done for the ground state properties of these isotopes within the Covariant Density Functional Theory with a Density-Dependent Point-Coupling X parametrization. The latter results, which are compared to indirect measurements of the $\beta_{2}$ deformations, serve as a guide in interpreting the most appropriate description offered by the two approaches involving the Bohr-Mottelson Hamiltonian with octic potential. After a thorough analysis of the numerical results in relation to the experimental data of these isotopes, an evident conclusion emerges. It is very hard to attribute only one of the two descriptions to the considered isotopes, while the complementary view of the data has proved to offer a more complete understanding of their structure of the states.
In other words, there are states of a band corresponding to a spherical vibrator symmetry, other states to a $\gamma$-unstable symmetry or to prolate one. Moreover, shape phase transition, shape coexistence and mixing between these three configurations are also present within this isotopic chain, respectively shape fluctuations and dynamical shape transitions. Another interesting behavior observed in these isotopes, is the staggering of the $\gamma$ band states which is well reproduced by the Octic $\&$ $\gamma$-unstable approach, while the Octic $\&$ prolate approach follows closely this grouping tendency of the states. If for the $\gamma$-unstable approach this is a typical signature, in the case of the Octic $\&$ prolate approach the deviation from the regular prolate behavior is related to the presence of the shape coexistence with mixing, respectively to the shape fluctuation phenomena.

\begin{acknowledgments}
This work was supported by grants of the Romanian Ministry of Research, Innovation and Digitalization, CNCS-UEFISCDI, project number PN-IV-P1-PCE-2023-0273, within PNCDI IV, and project number PN-23-21-01-01/2023, respectively through computational resources of HPC-MARWAN (www.marwan.ma) provided by the National Center for Scientific and Technical Research (CNRST) in Rabat, Morocco.
\end{acknowledgments}


\bibliography{apssamp}

\end{document}